\documentclass[useAMS,usenatbib,usegraphicx]{mn2e}

\usepackage{subfigure}
\usepackage{url}
\usepackage{comment}
\usepackage{color}
\usepackage{dcolumn}
\usepackage{enumitem}

\newcommand*{\rom}[1]{\uppercase\expandafter{\romannumeral #1}}
\newcommand\ion[2]{#1$\;${\small\rmfamily\rom{#2}}\relax}%

\newcommand\aap{{A\&A}}%
\newcommand\araa{{ARA\&A}}%
\newcommand\mnras{{MNRAS}}%
\newcommand\apj{{ApJ}}%
\newcommand\apjs{{ApJS}}%
\newcommand\apjl{{ApJ}}%
\newcommand\aj{{AJ}}%
\newcommand\pasp{{PASP}}%
\newcommand\spie{{\it Society of Photo-Optical Instrumentation
    Engineers (SPIE) Conference Series}}

\setlist[enumerate]{noitemsep}
\setlist[enumerate,1]{leftmargin=*}
\setlist[itemize]{noitemsep}
\setlist[itemize,1]{leftmargin=*}
\setlist[description]{noitemsep}
\setlist[description,1]{leftmargin=*}
\setenumerate[0]{label=(\arabic*)}

\makeatother

\title[Integrated CaT Measurements of M31 Clusters]{The Integrated
  Calcium II Triplet as a Metallicity Indicator: Comparisons with High
  Resolution [Fe/H] in M31 Globular Clusters}

\author[Sakari \& Wallerstein]{Charli
M. Sakari$^{1}$\thanks{E-mail: sakaricm@u.washington.edu} and George Wallerstein$^{1}$\\
$^{1}$ Department of Astronomy, University of Washington, Seattle WA
98195-1580, USA\\
}

\begin{document}

\maketitle

\label{firstpage}

\begin{abstract}
Medium resolution ($R = 4000-9000$) spectra of the near
infrared \ion{Ca}{2} lines (at 8498, 8542, and 8662 \AA) in M31
globular cluster integrated light spectra are presented.  In
individual stars the \ion{Ca}{2} triplet (CaT) traces stellar
metallicity; this paper compares integrated CaT strengths to well
determined, high precision [Fe/H] values from high resolution
integrated light spectra.  The target globular clusters cover a wide
range in metallicity (from $[\rm{Fe/H}] \sim -2.1$ to $-0.2$).  While
most are older than $\sim 10$ Gyr, some may be of intermediate age
(2-6 Gyr). A handful (3-6) have detailed abundances (e.g. low [Ca/Fe])
that indicate they may have been accreted from dwarf galaxies.  Using
various measurements and definitions of CaT strength, it is confirmed
that for GCs with $[\rm{Fe/H}]\la -0.4$ and older than $\sim 2$ Gyr the
integrated CaT traces cluster [Fe/H] to within $\sim 0.2$ dex,
independent of age.  CaT lines in metal rich GCs are very sensitive to
nearby atomic lines (and TiO molecular lines in the most metal rich
GCs), largely due to line blanketing in continuum regions.  The
[Ca/Fe] ratio has a mild effect on the integrated CaT strength in
metal poor GCs.  The integrated CaT can therefore be safely used to
determine rough metallicities for distant, unresolved clusters,
provided that attention is paid to the limits of the measurement
techniques.
\end{abstract}


\section{Introduction}\label{sec:Intro}
Galaxy formation is an active field of research in modern
astronomy.  Large surveys of nearby stars are working to untangle the
complex assembly history of the Milky Way by focusing on detailed
stellar abundances and kinematics.  These surveys aim to determine
when and where Milky Way stars have formed, establishing the
importance of satellite accretion, radial migration, etc.  A more
universal theory of galaxy formation, however, requires observations
of other types of galaxies (such as giant ellipticals) in different
environments (such as galaxy clusters).  Though the stellar
populations in nearby galaxies like M31 can be studied via photometry
and low to medium resolution spectroscopy, outside of the Local Group
individual stars are too faint and/or crowded to be easily resolved.
In lieu of detailed studies of individual stars, extragalactic studies
can utilize a galaxy's globular cluster (GC) population to probe the
properties of the host galaxy field stars.

Integrated photometry provides a relatively quick way to determine
information about a galaxy's GC population.  In particular, integrated
colours should correlate roughly with cluster metallicity, since metal
poor GCs are generally bluer than metal rich ones.  Integrated
photometric studies have demonstrated that nearly every massive galaxy
has a GC population that is bimodal in colour (see, e.g.,
\citealt{Peng2006})---if this reflects a metallicity bimodality, this
indicates that most galaxies host two chemically distinct GC
populations, which has major implications for galaxy formation
scenarios (see the review by \citealt{BrodieStrader2006}).  However,
the degenerate effects of age, chemical composition, and (most
notably) nonlinear colour-metallicity relationships
(e.g. \citealt{Yoon2006}) cast some doubt on the nature of the
observed colour bimodalities, and the presence of metallicity
bimodalities in most GC systems is still hotly debated.

The most direct way to determine metallicities of GCs is to turn to
integrated light (IL) spectroscopy, where a single spectrum is
obtained for an entire GC.  High resolution ($R\ga 20,000$) IL
spectroscopy resolves individual spectral lines, providing detailed
abundances for many elements---in particular, [\ion{Fe}{1}/H] can be
determined to high precision (with random uncertainties $<0.05$ dex,
e.g.,
\citealt{McWB,Colucci2009,Colucci2011a,Colucci2014,Sakari2013,Sakari2015}).
However, high resolution spectra require significant observing time,
even on 8 m class telescopes, and it is only feasible to obtain
detailed abundances for a handful of GCs per galaxy.  For instance,
\citet{Colucci2013} present abundances for 10 of the 1300 GCs in
NGC~5128 \citep{Harris2013} from $\sim100$ hours of observations on
the 6.5 m Magellan Clay Telescope.  While these studies
utilize the best possible methods for studying individual GCs, high
resolution IL spectroscopy is not suitable for science goals that
require metallicities of large samples of GCs (e.g. for studying
bimodalities, radial gradients, etc.) or for observations of objects
that have not yet been confirmed as GCs.  Such science goals require
determining GC metallicities from lower resolution spectra.  The Lick
indices (\citealt{Worthey1994}; most of which are in the blue) have
served as useful low resolution age- and metallicity-sensitive
features. Recently, however, attention has refocused on the near
infrared \ion{Ca}{2} triplet (CaT) index.

The CaT is a set of three strong, singly ionized calcium features at
8498,  8542, and 8662 \AA.  In individual stars the CaT features are
easily detectable at moderate spectral resolution ($R\sim 5000$) and
are primarily sensitive to metallicity
(e.g. \citealt{Battaglia2008,Starkenburg2010}).  Though
  the CaT lines are also sensitive to stellar surface gravity---an
  effect that renders the CaT lines useful for probing the initial
  mass function in early-type galaxies
  \citep{ConroyVanDokkum2012,Ferreras2013}--- well-populated GCs
  should not have significant population sampling differences, and the
  CaT should be primarily sensitive to metallicity.
The pioneering work of \citet[hereafter AZ88]{AZ88} established that
the integrated CaT strength is correlated with [Fe/H] in Milky Way
clusters.  More recently, \citet{Foster2010} obtained CaT spectra of
144 GCs associated with the early type galaxy NGC~1407; utilizing the
AZ88 CaT strength-[Fe/H] relation, they found several anomalies in the
behaviour of the integrated CaT.  In particular, the brightest
NGC~1407 GCs had similar CaT strengths, despite large colour
differences.  Furthermore, the clear colour bimodality in the GC
population was not seen in CaT strength.\footnote{\citet{Foster2010}
  did see bimodality in the CaT strengths in their template-fitted
  spectra (see Section \ref{subsubsec:Templates}), but the shape of
  the distribution is different from the shape of the colour
  distribution.  They did not see any bimodality in CaT strength when
  the features were measured on the observed spectra.}  Foster et
al. also found that the predicted relationship between CaT strength
and [Fe/H] varied significantly between stellar population models.
Their explanations for these discrepancies included unknown variations
in the GC populations (e.g. as a result of age, horizontal branch
morphology, etc.) or changes in the CaT strength/metallicity
relationship at high metallicities.  Without an understanding of how
these effects could alter the integrated CaT, Foster et al. noted that
CaT studies of unresolved populations could remain ``problematic.''

\citet{Usher2012} then utilized single stellar population (SSP) models
to derive a new relationship between CaT strength and total
metallicity, [Z/H].  With this relation, they derived CaT
metallicities for 903 GCs from 11 early type galaxies.  Comparing with
integrated photometry, they find galaxy to galaxy differences in their
colour-CaT metallicity relations, which they argue may be due to age
or initial mass function (IMF) differences between GC systems.
The individual GC spectra from each galaxy were then grouped by colour
and stacked together by \citet{Usher2015}.  This stacking process
improved S/N ratios for similar GC spectra, and supports their earlier
findings that the CaT strength-colour relationship varies between
galaxies.  Usher et al. suggested that this could be due to variations
in GC age or detailed abundances.

These recent papers have only presented indirect evidence that the IL
CaT tracks GC metallicity, through comparisons with colours, SSP
models, and Lick index metallicities (which are also often derived
with SSPs).  Of course, \citet{Foster2010} and \citet{Usher2012} could
not compare to high resolution [Fe/H] ratios, because none are yet
available for such faint clusters.  This paper presents the first
direct comparison between integrated CaT strengths and high resolution
metallicities since AZ88, through observations of M31 GCs.  For this
type of comparison, M31 GCs are preferable to Milky Way GCs, even
though there is more information available for Milky Way GCs, for two
reasons.
\begin{enumerate}
\item IL spectra are easier to obtain for M31 GCs.  Milky Way GCs are
  nearby, and obtaining complete IL spectra requires scanning across
  the clusters out to the half-light radii.  For optical spectral
  lines incomplete IL spectra (e.g. of only GC cores) can introduce
  uncertainties in [Fe/H] of up to 0.1-0.2 dex as a result of mass
  segregation and stochastic sampling \citep{McWB,Sakari2014}.  M31
  GCs are sufficiently small ($r_h \la 4\arcsec$; \citealt{RBCref}) to
  obtain a complete IL spectrum in a single pointing.  This also
  ensures that the CaT strengths and high resolution [Fe/H] ratios are
  from the same stellar populations.
\item There are more GCs in M31 than in the Milky Way; similarly, M31
  GCs cover regions in parameter space that the Milky Way GCs do not.
  In particular, M31 has bright clusters that extend to higher
  velocity dispersions, younger ages, and lower [$\alpha$/Fe] ratios.
  Since the CaT calibration on MW GCs has already been done by AZ88,
  it is essential to test the calibration on GCs that are unlike
  typical MW GCs.
\end{enumerate}

The goal of this paper is to investigate the validity of the
integrated CaT as a metallicity indicator by directly comparing CaT
strengths to high resolution IL abundances, {\it without adopting any
  SSP models}.  The M31 GC CaT data are described in Section
\ref{sec:Observations}.  CaT measurement methods are discussed and
explored in Section \ref{sec:Measurements}.  With the best
measurements of the CaT lines, the trends with [Fe/H] are explored in
Section \ref{sec:Metallicity}.  The feasibility of CaT studies
for unresolved, extragalactic GCs is then discussed in Section
\ref{sec:Discussion}.

\section{CaT Observations and Data Reduction}\label{sec:Observations}
The CaT spectra were obtained at Apache Point Observatory (APO) and
Kitt Peak National Observatory (KPNO) in the Fall of 2014.  Targets
were selected from the high resolution samples of \citet{Colucci2014},
\citet{Sakari2015}, and Sakari et al. (2016, {\it in prep.});
priority was placed on clusters that were bright, covered a wide
metallicity range, and had unusual ages or [$\alpha$/Fe] ratios.  The
details of the targets are shown in Table \ref{table:Targets}. 

\begin{table*}
\centering
\begin{minipage}{165mm}
\begin{center}
\caption{Target information.\label{table:Targets}}
  \begin{tabular}{@{}lccccccccccc@{}}
  \hline
   &  & & & & & & \multicolumn{5}{l}{Literature High Resolution Values}\\
Cluster & RA (hms) & Dec (dms) & $V_{\rm{int}}$ & $t_{\rm{exp}}$ & S/N$^{a}$ & $v_{\rm{helio}}^{b}$ & $\sigma$ &[Fe/H] & Age & [Ca/Fe] & Refs\\
   & J2000 & J2000 &  & (sec) &  & (km s$^{-1}$) & (km s$^{-1}$) &  & (Gyr) & & \\
\hline
B006      & 00:40:26.5 & $+$41:27:26.4 & 15.5 & 1800 & 140 & $-227$ & 10.12 & -0.83 & 12.0 & 0.25 & 1\\
B012      & 00:40:32.5 & $+$41:21:44.2 & 15.0 & 1200 & 80 & $-382$ & 19.50 & -1.61 & 11.5 & 0.40 & 2\\
B029      & 00:41:17.8 & $+$41:00:22.8 & 16.6 & 5400 & 70 & $-485^{d}$ & 6.51  & -0.43 & 2.1  & 0.04 & 2\\
B045$^{c}$ & 00:41:43.1 & $+$41:34:20.0 & 15.8 & 4800 & 160 & $-442$ & 10.24 & -0.94 & 12.5 & 0.22 & 2\\
B063      & 00:42:00.9 & $+$41:29:09.5 & 15.7 & 2700 & 190 & $-295$ & 14.81 & -1.10 & 14.0 & 0.36 & 1\\
B088      & 00:42:21.1 & $+$41:32:14.3 & 15.0 & 1200 & 105 & $-469$ & 14.25 & -1.71 & 14.0 & 0.21 & 2\\
B110      & 00:42:33.1 & $+$41:03:28.4 & 15.3 & 1800 & 90 & $-222$ & 18.20 & -0.68 & 6.5  & 0.14 & 2\\
B163      & 00:43:17.0 & $+$41:27:44.9 & 15.0 & 1200 & 100 & $-189$ & 17.41 & -0.49 & 11.5 & 0.28 & 2\\
B171      & 00:43:25.0 & $+$41:15:37.1 & 15.3 & 2400 & 200& $-264$ & 16.86 & -0.45 & 12.5 & 0.27 & 1\\
B182      & 00:43:36.7 & $+$41:08:12.2 & 15.4 & 1800 & 140 & $-377$ & 19.29 & -1.04 & 12.5 & 0.41 & 2\\
B193      & 00:43:45.5 & $+$41:36:57.5 & 15.3 & 1800 & 160 & $-80$  & 15.79 & -0.16 & 8.0  & 0.17 & 2\\
B225      & 00:44:29.8 & $+$41:21:36.6 & 14.2 & 1200 & 220& $-141$ & 25.73 & -0.66 & 10.0 & 0.40 & 2\\
B232$^{c}$ & 00:44:40.5 & $+$41:15:01.4 & 15.7 & 3535 & 150 & $-201$ & 14.24 & -1.77 & 14.0 & 0.34 & 2\\
B240      & 00:45:25.2 & $+$41:06:23.8 & 15.2 & 1800 & 70 & $-44$  & 12.23 & -1.54 & 12.5 & 0.30 & 2\\
B311      & 00:39:33.8 & $+$40:31:14.4 & 15.5 & 1800 & 60 & $-527$ & 12.77 & -1.71 & 14.0 & 0.31 & 1\\
B381$^{c}$ & 00:46:06.6 & $+$41:20:58.9 & 15.8 & 3600 & 105 & $-91$  & 9.87  & -1.17 & 12.5 & 0.27 & 2\\
B383      & 00:46:12.0 & $+$41:19:43.2 & 15.3 & 1800 & 80 & $-208$ & 11.13 & -0.78 & 12.5 & 0.28 & 2\\
B384      & 00:46:21.9 & $+$40:17:00.0 & 15.8 & 4200 & 135 & $-379$ & 9.00  & -0.63 & 6.5  & 0.14 & 2\\
B386      & 00:46:27.0 & $+$42:01:52.8 & 15.6 & 3300 & 105 & $-391$ & 11.42 & -1.14 & 11.0 & 0.27 & 2\\
B405      & 00:49:39.8 & $+$41:35:29.7 & 15.2 & 1200 & 120 & $-180$ & 12.29 & -1.33 & 12.5 & 0.26 & 2\\
B457      & 00:41:29.0 & $+$42:18:37.7 & 16.9 & 11700& 55 & $-346^{e}$ & 4.73  & -1.23 & 11.0 & 0.08 & 2\\
B472      & 00:43:48.4 & $+$41:26:53.0 & 15.2 & 1200 & 80 & $-113$ & 14.37 & -1.16 & 10.0 & 0.32 & 1\\
B514$^{c}$ & 00:31:09.8 & $+$37:53:59.6 & 15.8 & 3600 & 100 & $-489$ & 8.49  & -1.74 & 14.0 & 0.45 & 2\\
G002      & 00:33:33.8 & $+$39:31:18.5 & 15.9 & 2760 & 105 & $-319$ & 10.12 & -1.63 & 11.5 & -0.02 & 2\\
H10       & 00:35:59.7 & $+$35:41:03.6 & 15.7 & 4800 & 55 & $-322$ & 6.60  & -1.36 & 12.0 & 0.25 & 3\\
H23       & 00:54:25.0 & $+$39:42:55.5 & 16.8 & 5400 & 70 & $-381$ & 6.20 & -1.12 & 9.0 & 0.41 & 3\\
MGC1$^{c}$ & 00:50:42.5 & $+$32:54:58.7 & 15.5 & 4800 & 70 & $-370$ & 8.29 & -1.56 & 11.5 & 0.18 & 2\\
PA06      & 00:06:12.0 & $+$41:41:21.0 & 16.5 & 9600 & 80 & $-321$ &  5.60 & -2.06 & 12.0 & 0.46 & 3\\
PA17      & 00:26:52.2 & $+$38:44:58.1 & 16.3 & 6000 & 90 & $-242$ &  6.10 & -0.93 & 12.0 & 0.04 & 3\\
PA53      & 01:17:58.4 & $+$39:14:53.2 & 15.4 & 1800 & 100 & $-255$ & 12.00 & -1.64 & 12.0 & 0.19 & 3\\
PA54      & 01:18:00.1 & $+$39:16:59.9 & 15.9 & 2640 & 70 & $-329$ &  7.50 & -1.84 & 13.0 & 0.28 & 3\\
PA56      & 01:23:03.5 & $+$41:55:11.0 & 16.8 & 10800& 45 & $-226$ &  6.40 & -1.73 & 12.0 & 0.24 & 3\\
\hline
\end{tabular}
\end{center}
\end{minipage}\\
\medskip
\raggedright {\bf References: } Positions and magnitudes are from the
Revised Bologna Catalog (RBC; \citealt{RBCref}) and \citet{Huxor2014}.
Velocity dispersions, ages, and abundances are from the high
resolution studies of 1) Sakari et al. (2016, {\it in prep.}), 2)
\citet{Colucci2014}, and 3) \citet{Sakari2015}.\\
$^{a}$ S/N ratios are determined at 8500 \AA \hspace{0.02in} and are
per resolution element.\\
$^{b}$ Typical uncertainties in the heliocentric radial
  velocity are $1-5$ km s$^{-1}$.\\
$^{c}$ Target was observed at KPNO.\\
$^{d}$ This radial velocity does not agree with the RBC value
(by $\sim100$ km s$^{-1}$) but does
agree with \citet{Colucci2014} and \citet{Caldwell2011}.\\
$^{e}$ This radial velocity does not agree with the RBC or
\citet{Colucci2014}, but does agree with \citet{Caldwell2011}.
      The discrepancy with Colucci et al. is $\sim 300$ km s$^{-1}$.
Because of the disagreement with the radial velocity from the high
resolution study, this target has been removed from the metallicity
calibration (though it is retained for the measurement tests in
Section \ref{sec:Measurements}).\\ 
\end{table*}

\subsection{Observations: APO}\label{subsec:APOObservations}
CaT spectra of 27 clusters were obtained with the Dual Imaging
Spectrograph (DIS) on the 3.5 m telescope at APO in Fall 2014.  The
R1200 grating was used in combination with the 1.\arcsec5 slit,
yielding a spectral resolution of 0.56 \AA/pix (or $R\sim 4,000$ at
8500 \AA) and wavelength coverage from $\sim8000 - 9100$ \AA.  The
slit is 6$\arcmin$ long, providing full coverage of the clusters
past their half-light radii and allowing simultaneous sky observations
on either side of the cluster---sky observations are essential because
the CaT region is heavily contaminated by strong atmospheric OH
emission lines (see Fig. 2 in \citealt{Battaglia2008}).  Exposure
times were calculated to achieve S/N ratios of 100, though bad weather
and poor seeing meant that some clusters did not have receive
sufficient time to reach this goal.  Most of the clusters have
excellent S/N ratios, as demonstrated in Table \ref{table:Targets}.
The largest spectral contamination comes from the strong night sky
emission lines.

\subsection{Observations: KPNO}\label{subsec:KPNOObservations} 
Spectra of five additional GCs were obtained with the WIYN (Wisconsin,
Indiana, Yale, and NOAO) 3.5 m telescope with the Sparsepak IFU
\citep{Bershady1,Bershady2} in single object mode feeding the Bench
Spectrograph \citep{Bench1,Bench2}.  Sparsepak provides 82 fibres
sparsely arranged over a $72\arcsec \times 71 \arcsec$ region; each
fibre has a diameter of $4.\arcsec7$ \citep{Bershady1}, which covers a
single GC past its typical half-light radius.  In single object mode
only the central fibre captures GC light; two separate fibres on the
edges are then utilized for sky observations.  The Bench
Spectrograph's ``echelle'' grating (316 gr/mm, blazed at
63.4$^{\circ}$) provides a slightly higher spectral resolution
($\sim~0.28$ \AA/pix, or $R\sim 9700$) than the APO spectra, but a
smaller wavelength coverage from $\sim 8300-8820$ \AA.  As with the
APO targets, sky line contamination is the biggest issue with these
spectra.

\subsection{Stellar Templates}\label{subsec:Templates}
Stellar sources were also observed to serve as templates for spectral
fitting (see Section \ref{subsubsec:Templates}).  These templates were
chosen to span a range in metallicity (though there are no
$\alpha$-deficient, metal poor stellar templates), temperature, and
luminosity class (though most of the templates are cool giants, since
IL spectra are dominated by red giant branch stars). All template
targets are stars that could be found in GCs, have magnitudes
$V=8-10$, and were observable in minutes---they therefore have high
S/N ratios and are uncontaminated by sky emission lines. Stellar
targets were observed with both telescopes, in order to match spectral
resolution; 16 template stars were observed at APO and only 10 at
KPNO.

\subsection{Data Reduction}\label{subsec:DataReduction}
The data reduction was performed in the Image Reduction and
Analysis Facility program (IRAF).\footnote{IRAF is distributed by the
  National Optical Astronomy Observatory, which is operated by the
  Association of Universities for Research in Astronomy, Inc., under
  cooperative agreement with the National Science Foundation.}
The DIS data were reduced following the standard procedures for long
slit spectra, while the Sparsepak data were reduced with the {\it
  dohydra} task.  The spectra were extracted with variance weighting,
though the continuum levels of the non-variance-weighted spectra were
maintained (see the discussion in \citealt{Sakari2013} for high
resolution spectra).  Sky subtraction was performed after aligning the
sky lines, since the curved projection of the slit onto the detector
\citep{Minkowski1942} leads to small offsets between the target
and sky spectra.  Continuum levels were fit with low order
polynomials across the entire observed wavelength range, and the
spectra were roughly normalized.  This initial continuum normalization
was done in order to combine individual observations using
sigma-clipping procedures.  The implications of continuum
normalization will be discussed in more detail in Sections
\ref{sec:Measurements} and \ref{sec:Metallicity}.

Radial velocities were determined through cross-correlations with a
medium resolution spectrum of Arcturus, where the high resolution, high
S/N Arcturus spectrum from
\citet{Hinkle2003}\footnote{\url{ftp://ftp.noao.edu/catalogs/arcturusatlas/}}
was downgraded to match the DIS resolution.  The final, heliocentric
radial velocities are shown in Table \ref{table:Targets}.  For the
most part the radial velocities match those from \citet{Colucci2014},
\citet{Caldwell2011}, and the Revised Bologna Catalog
(RBC;
\citealt{RBCref,RBCref2}),\footnote{\url{http://www.bo.astro.it/M31/}}
with the exception of two clusters: B029 and B457.  The B029 radial
velocity disagrees with the value in the RBC (by $\sim 100
 $ km s$^{-1}$), while the B457 velocity
disagrees with the RBC and with \citet{Colucci2014} (by
  $\sim 300$ km s$^{-1}$); both GCs are in
agreement with the values in \citet{Caldwell2011}.  Because B457's
radial velocity disagrees with the high resolution study, it was
removed from the [Fe/H] calibration in Section \ref{sec:Metallicity}.

After the spectra were shifted to the rest frame, individual exposures
were combined with average sigma-clipping routines, weighted by flux.
Limits were set to eliminate the effects of the sky lines, which are
especially prevalent around the reddest CaT line.

\section{Measuring CaT Strength}\label{sec:Measurements}
There are two important aspects involved in measuring the strengths of
the CaT lines.
\begin{enumerate}
\item Continuum identification.
\item The method for measuring line strength.
\end{enumerate}
Although the combined spectra are roughly normalized during the data
reduction procedure, this is likely to be insufficient for measuring
the strengths of the CaT lines. The techniques for measuring the lines
may also lead to systematic differences in their strengths.

\subsection{Continuum Identification}\label{subsec:Bandpass}
To determine the best methods for identifying the continuum levels,
simple integrations of the line profiles are utilized---referred to as
``pseudo equivalent widths'' by AZ88, these measurements have also
known as ``indices.''  To measure the pseudo EWs
in the M31 GCs, the {\tt indexf} program \citep{Cardiel2010} was used;
this code also determines formal errors in the measured indices
\citep[hereafter C01]{Cenarro2001}.  To avoid offsets from sky lines
and/or low S/N, the template-fitted spectra from Section
\ref{subsubsec:Templates} are used.  Three continuum definitions are
explored.
\begin{enumerate}
\item The definitions of AZ88, which are determined individual for
  each CaT line based on defined regions on other side of each line.
\item The CaT index definitions of C01, which are determined across all
  three lines, again using defined regions around the CaT features.
  These definitions are more restrictive than the AZ88 definitions,
  and are designed to avoid known atomic lines.
\item Averaged continuum regions around each spectral line, with known
  atomic features and sky line residuals masked out.  As with the AZ88
  definitions, these regions are determined around each CaT line.
\end{enumerate}

\begin{figure*}
\begin{center}
\centering
\subfigure{\includegraphics[scale=0.55,trim=1.25in 0in 0.05in 0.0in]{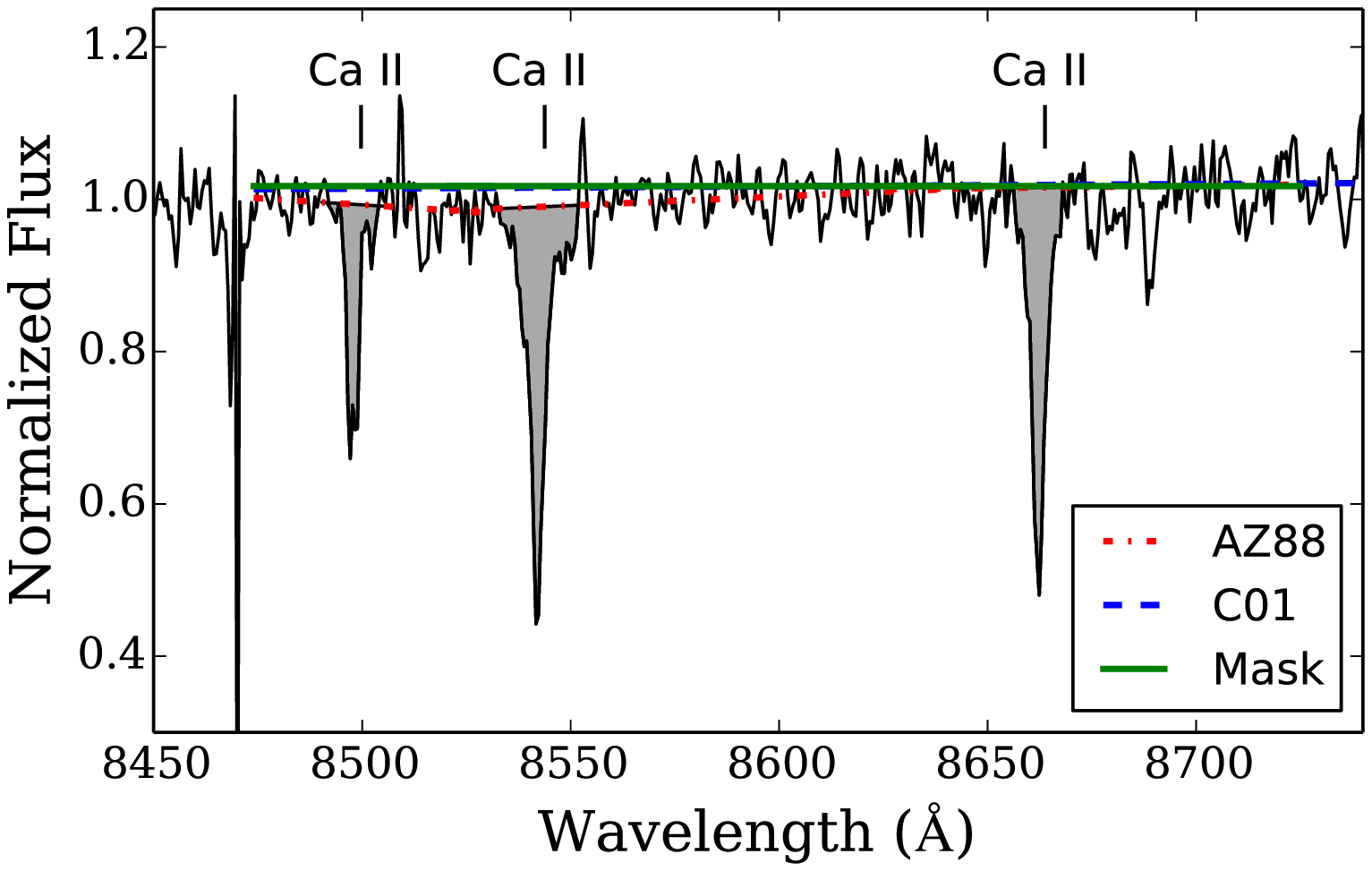}\label{fig:B163_C01}}
\subfigure{\includegraphics[scale=0.55,trim=0.5in 0in 1.25in 0.0in]{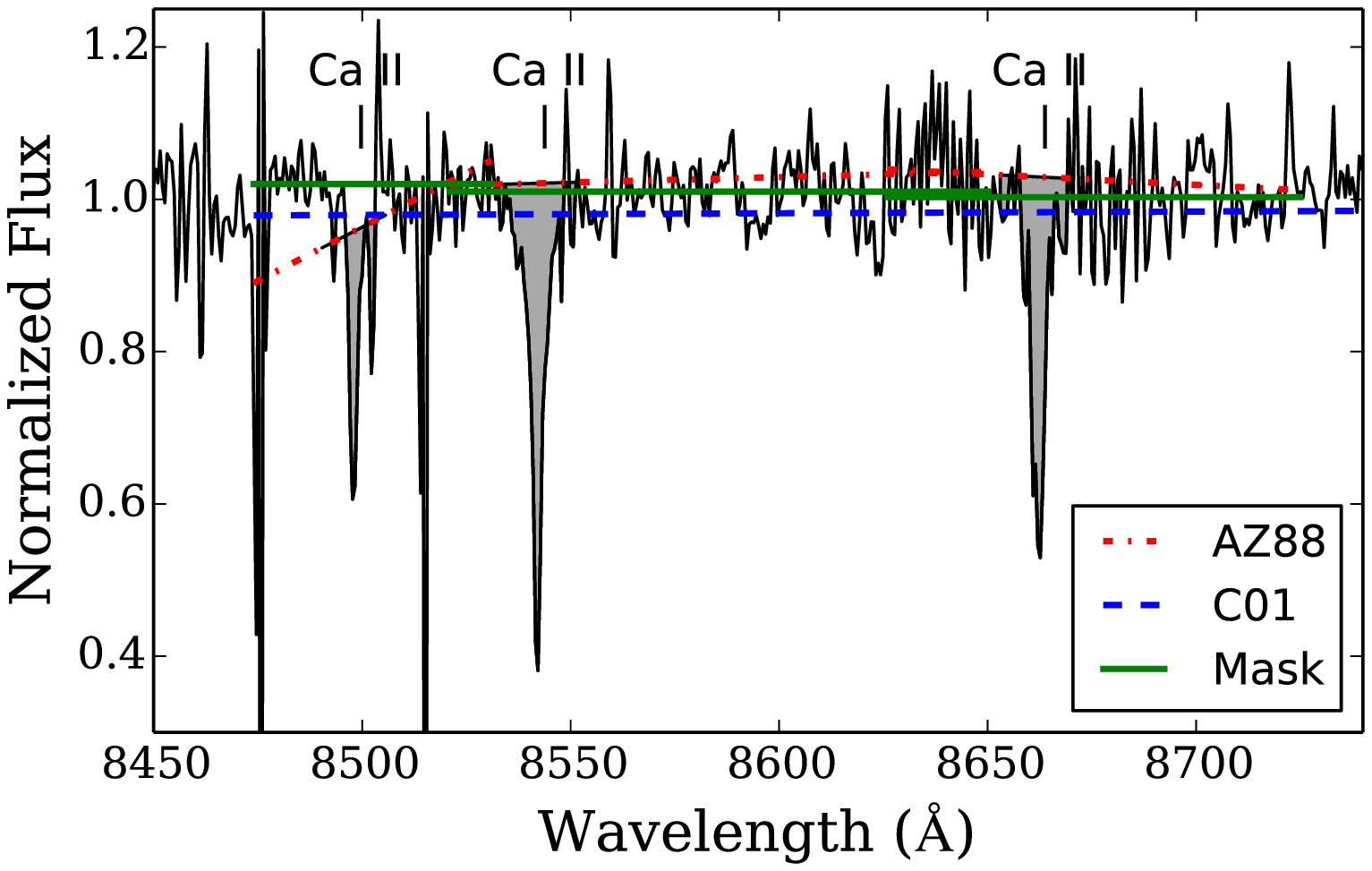}\label{fig:H23_C01}}
\caption{Continuum fits with the AZ88 (dot-dashed red lines) and C01
  continuum definitions (dashed blue lines) compared to the continuum
  fitting with masked out regions (sold green lines).  Two GCs are
  shown: the metal rich GC B163 (left) and the lower S/N, more metal
  poor GC, H23 (right).  Note that in the left figure the C01 continuum
  definition is identical to the masked fits.  The C01 definitions and
  the masked regions provide better estimates of the continuum for
  metal rich GCs, though the C01 definitions may be more sensitive to
  noise and sky lines in lower quality spectra.}\label{fig:C01}
\end{center}
\end{figure*}

\begin{figure*}
\begin{center}
\centering
\subfigure{\includegraphics[scale=0.55,trim=1.25in 0in 0.05in 0.0in]{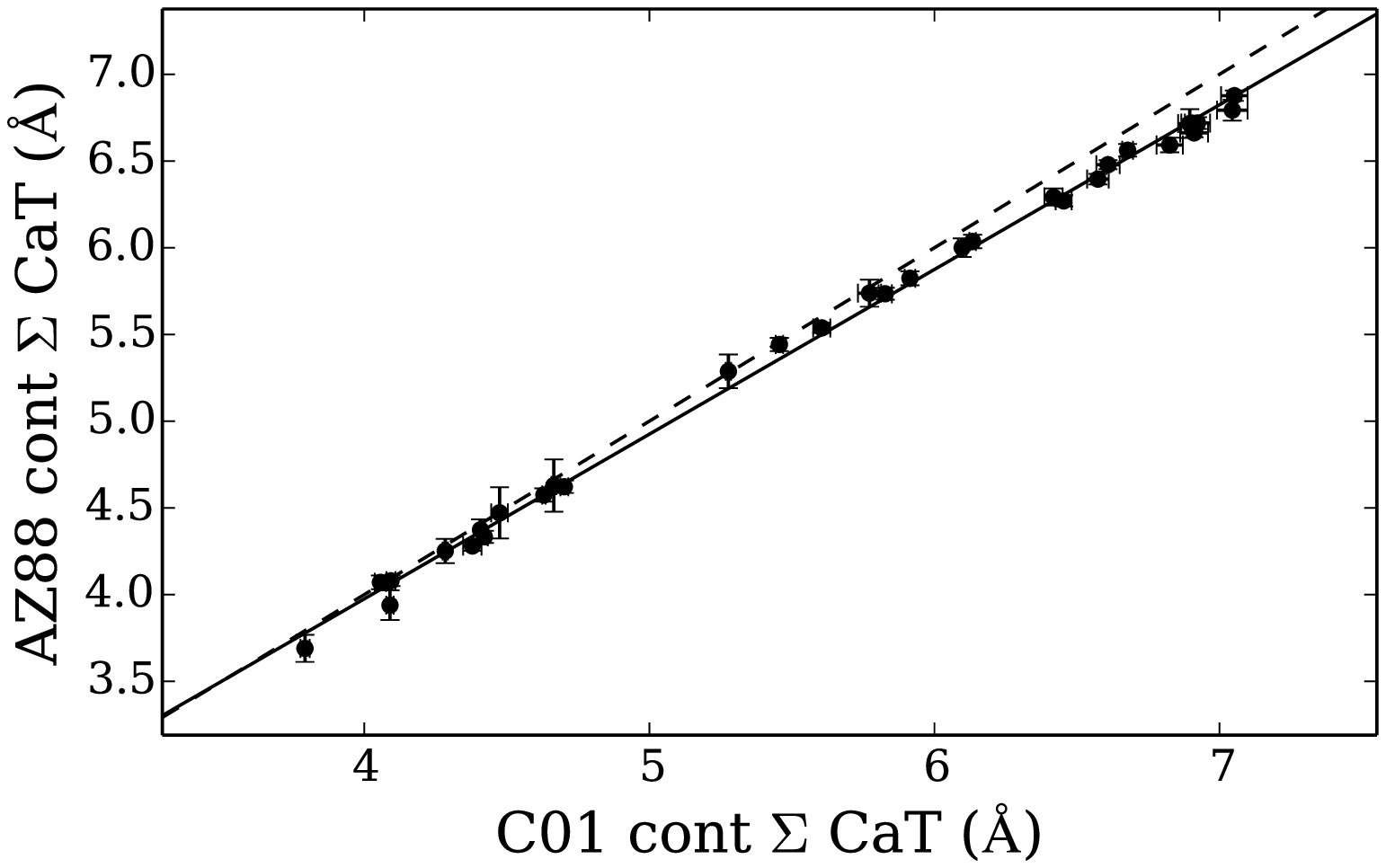}}
\subfigure{\includegraphics[scale=0.55,trim=0.5in 0in 1.25in 0.0in]{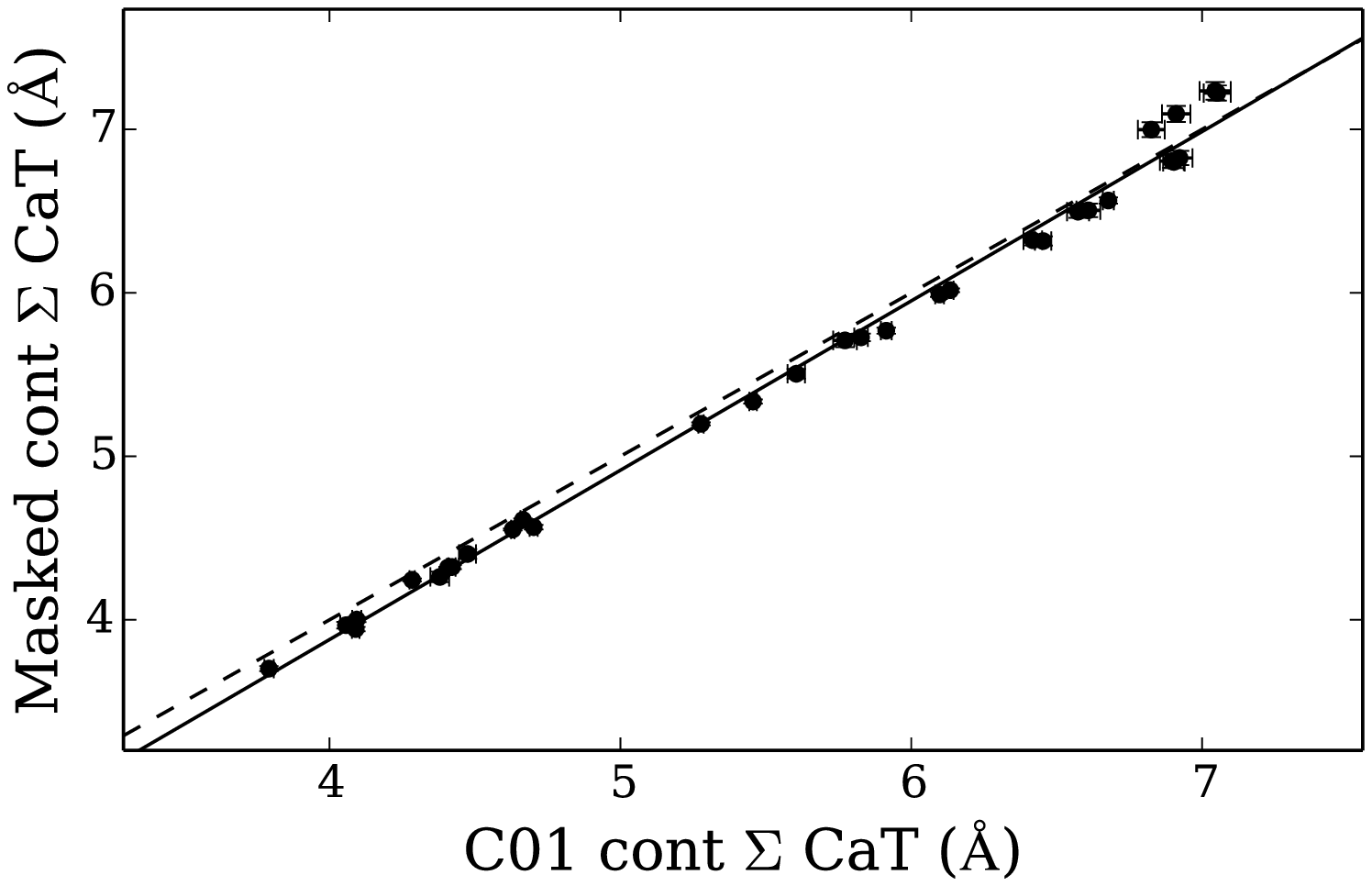}\label{fig:C01CompVoigt}}
\caption{Comparisons between measurements of the IL CaT strength in
  the template-fitted spectra with different continuum definitions.
  In every case the line definitions of AZ88 are adopted, with varying
  continuum fits.  {\it Left: } The AZ88 vs. C01 continuum
  definitions.  {\it Right:} Averaged continuum fits to the observed data with
  sky lines and known atomic lines masked out, versus the C01
  continuum definitions.  The points are the M31 GCs.  The dashed line
  shows equal values, while the solid black line shows a least squares
  fit.
  \label{fig:C01Comp1}}
\end{center}
\end{figure*}

Figure \ref{fig:C01} shows sample CaT spectra with various continuum
fits.  The line measurements themselves are compared in Figure
\ref{fig:C01Comp1}.  The AZ88 definitions lead to smaller measurements
at strong CaT strengths (i.e. at high metallicities), while the C01
and masked continua are fairly similar except at the highest
metallicities. This can be understood by examining the continuum
definitions themselves.  C01 provide a comparison of several popular
indices, illustrating that most definitions are unsuitable for very
early or very late type stars.  In particular, spectra of hot stars
are affected by strong Paschen lines, while those of cool M stars are
dominated by TiO bands (see Figures 1-3 in C01). IL spectra are
dominated by RGB stars, particularly in the $I$ band. However,
contributions from cool M giants may become non-negligible in metal
rich GCs.  Additionally, the other definitions may not be suitable
over a wide range in [Fe/H], S/N, velocity dispersion, etc. The C01
indices are designed to be ``generic,'' i.e. applicable to stars of a
range of spectral types.  Furthermore, the narrower definitions for
the continuum regions better avoid significant atomic features which
are present in metal rich spectra.

The advantage of the C01 CaT continuum definition is demonstrated in
Figure \ref{fig:B163_C01}, which shows the C01 and AZ88 continuum fits
to the metal rich GC B163.  The AZ88 continuum is fit across several
strong atomic lines, which lowers the effective continuum.  The
narrower C01 definitions do a better job fitting the continuum level
in this metal rich GC (though they might still be affected by line
blanketing from atomic and/or molecular features).  The potential
disadvantages of the C01 continuum definitions are evident in Figure
\ref{fig:H23_C01}, where the lower S/N GC H23 has strong sky line
residuals.  The AZ88 continuum definitions work well except for the
bluest CaT line, CaT1, where some noise has strongly affected the
continuum level.  The C01 continuum is fit across the whole spectrum,
and is systematically lowered by sky lines and noise. This problem can
be solved by utilizing error spectra in the {\tt   indexf} program;
however, this is unnecessary for high S/N spectra. With
template-fitted spectra, the C01 continuum definitions or continuum
fits that mask out sky lines or atomic lines will be superior to the
AZ88 definitions across a wide metallicity range. 

These tests therefore demonstrate that the AZ88 continuum definitions
are unsuitable for metal rich GCs.  The C01 definitions perform
well, except in the case of strong sky lines.  The more rigorous
method of fitting the continuum with known sky and atomic lines masked
out can overcome this problem, as can the use of template-fitted
spectra.

\subsection{Line Strengths}\label{subsec:Measurements}
Measuring the strength of the CaT lines is a nontrivial process
because the lines are strong (i.e. saturated) and may be blended with
other features.  As discussed earlier, the original AZ88 calibration
utilized ``pseudo equivalent widths'' (EWs), which measure the area
within a defined region.  Individual stellar analyses
(e.g. \citealt{Battaglia2008}) often fit Gaussian profiles to the
spectral lines---however, the CaT lines have strong Lorentzian wings
and are distinctly non-Gaussian. \citet{Battaglia2008} compensate for
the wings by adding some factor to their Gaussian EWs; however the
lines can also be fit with Voigt profiles.  All these options are
explored to identify the ideal option for the current calibration and
for future extragalactic studies.

The extragalactic analyses of \citet{Foster2010} and \citet{Usher2012}
use template fits to their (often low S/N) spectra.  Although template
fits are not strictly necessary for these M31 GCs, they do help reduce
scatter in plots; the observed spectra and the template fits are
therefore considered for the tests below.

\subsubsection{Template fits}\label{subsubsec:Templates}
Though the CaT lines in the observed M31 spectra can be easily
measured, this is not feasible for more distant systems where the S/N
is much lower.  \citet{Foster2010} and \citet{Usher2012} combat this
problem by fitting the observed spectra with a combination of stellar
templates, which include 11 giants and 2 dwarfs of a range of
metallicities and temperatures.  For the APO targets sixteen stellar
templates are utilized,  encompassing a wide range in metallicity
(from [Fe/H]$\sim -2$ to $0$), surface gravity, and effective
temperature; for the KPNO targets only ten stellar templates were
observed, with a similar range in parameters.  All the template stars
were selected to be stars that could be found in GCs; in particular,
several horizontal branch stars were observed, including a very blue
one.

\begin{figure*}
\begin{center}
\centering
\subfigure{\includegraphics[scale=0.55,trim=1.25in 0in 0.05in 0.0in]{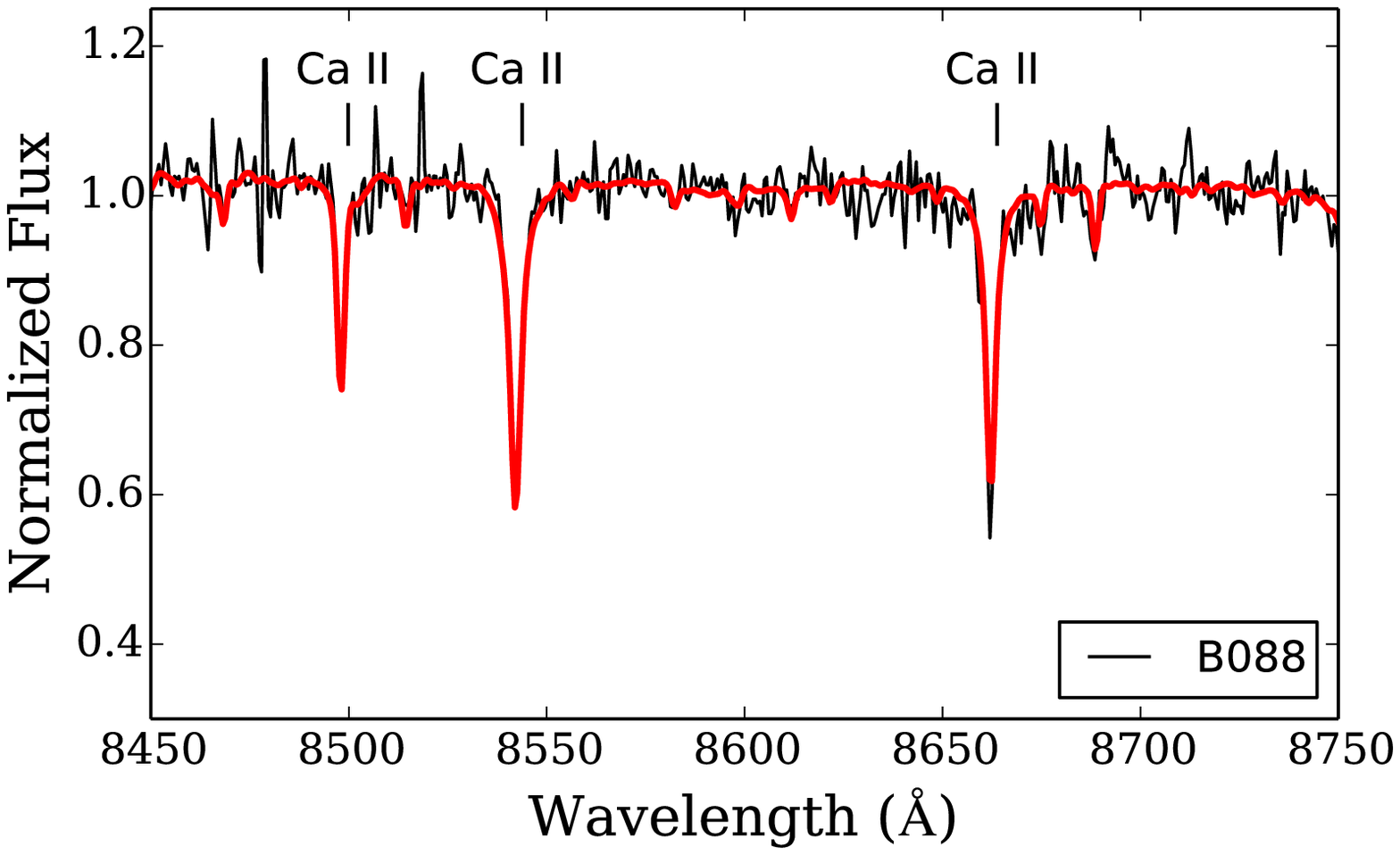}\label{fig:B088Template}}
\subfigure{\includegraphics[scale=0.55,trim=0.5in 0in 1.25in 0.0in]{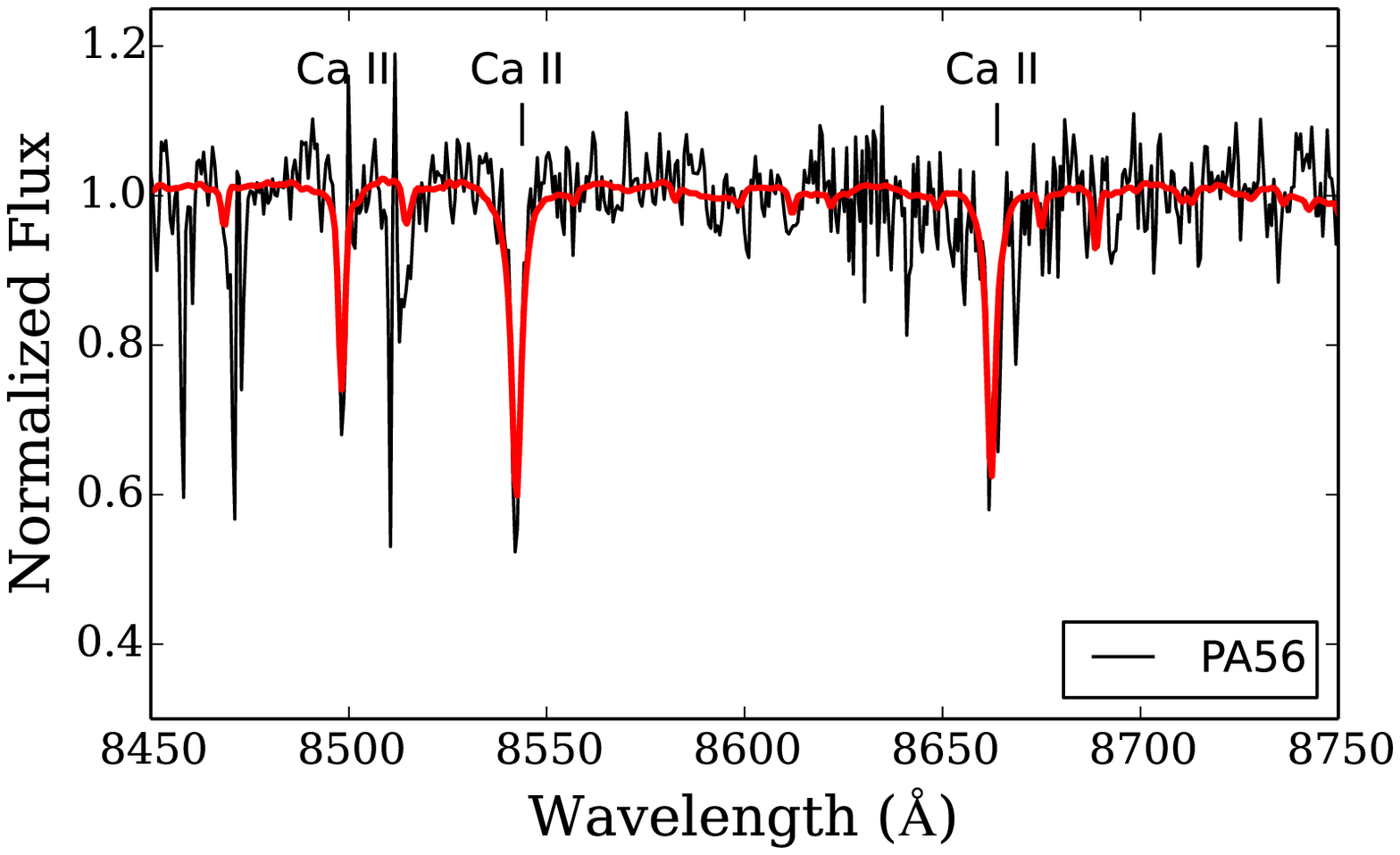}\label{fig:PA56Template}}
\caption{Template fits (red) to the observed B088 and PA56 spectra
  (black).  The three CaT lines are indicated.\label{fig:TemplateFits}}
\end{center}
\end{figure*}

As in \citet{Foster2010} and \citet{Usher2012}, the penalized
pixel-fitting (pPXF) code of \citet{pPXFref} was used to find the
linear combination of stellar templates that best fits a given
observed spectrum.  The templates were fit over the $8440-8820$
\AA \hspace{0.025in} region for the KPNO/WIYN spectra and the
$8440-8950$ \AA \hspace{0.025in} region for the DIS spectra.  Sample
template fits to B088 and PA56 are shown in Figure
\ref{fig:TemplateFits}.  Even lower S/N clusters with strong residual
sky line contamination (such as PA56) are reasonably well fit by the
templates.  Several clusters, including B088, are best fit with a
small contribution from the hottest blue horizontal branch star, which
leads to very weak Paschen lines in the fitted spectra (which was also
seen in the stacked spectra from \citealt{Usher2015}). These Paschen
lines do not have a strong effect on the CaT line strengths.  Errors
in the template fits were estimated with Monte Carlo resampling of the
observed spectra one hundred times.

\subsubsection{Line Integrations in Observed Spectra}\label{subsubsec:Integrals}
\begin{figure}
\begin{center}
\centering
\includegraphics[scale=0.5]{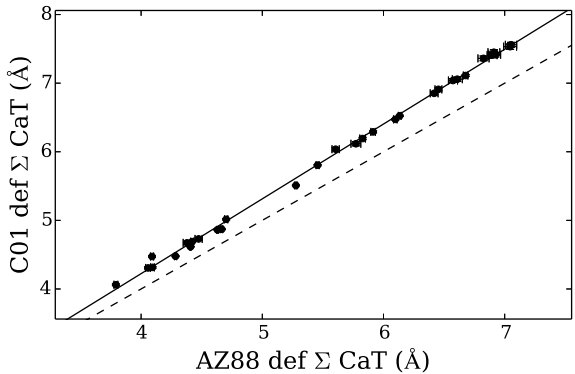}
\caption{Measurements with the C01 line definitions compared to the
  AZ88 definitions.  Both measurements utilize the C01 continuum
  regions.}\label{fig:LineDef}
\end{center}
\end{figure}

Again, the {\tt indexf} program \citep{Cardiel2010} was used to
measure pseudo EWs and their associated errors.  Given the results in
Section \ref{subsec:Bandpass} the C01 continuum definitions are
adopted.  The AZ88 and C01 line definitions are considered; the C01
definitions are slightly wider than the AZ88 ones, and are likely to
cover the wings of the lines more fully (especially for GCs with large
velocity dispersions). This measurement technique works well for
spectra of bright targets that are not significantly affected by sky
line residuals. Many of these M31 GC spectra, however, are affected by
sky lines, which will a) systematically affect the measurements of the
CaT lines and b) add scatter to any correlation with [Fe/H].  These
effects  are undesirable for a test of the correlation with [Fe/H],
and comparisons are only performed on the template-fitted
spectra. Figure \ref{fig:LineDef} compares measurements with the C01
and AZ88 definitions, both using the C01 continuum definitions.
Because the C01 indices are wider  than the AZ88 indices they are
stronger, with the offset increasing with CaT strength.  The C01 line
definitions are utilized in all subsequent integrations; note that the
fit in Figure \ref{fig:LineDef} can be used to convert AZ88
definitions to C01 definitions.

\subsubsection{Gaussian and Voigt Fits}\label{subsubsec:Gaussians}
Given noise and possible sky line contamination in the line profiles,
it will be more accurate to fit full spectral line profiles than to
simply integrate within defined regions.  CaT lines are often measured
with Gaussians (e.g. \citealt{Battaglia2008}), but as the CaT lines
are very strong, Voigt profiles are more appropriate.

\begin{figure*}
\begin{center}
\centering
\subfigure{\includegraphics[scale=0.45,trim=1.25in 0in 0.05in 0.0in]{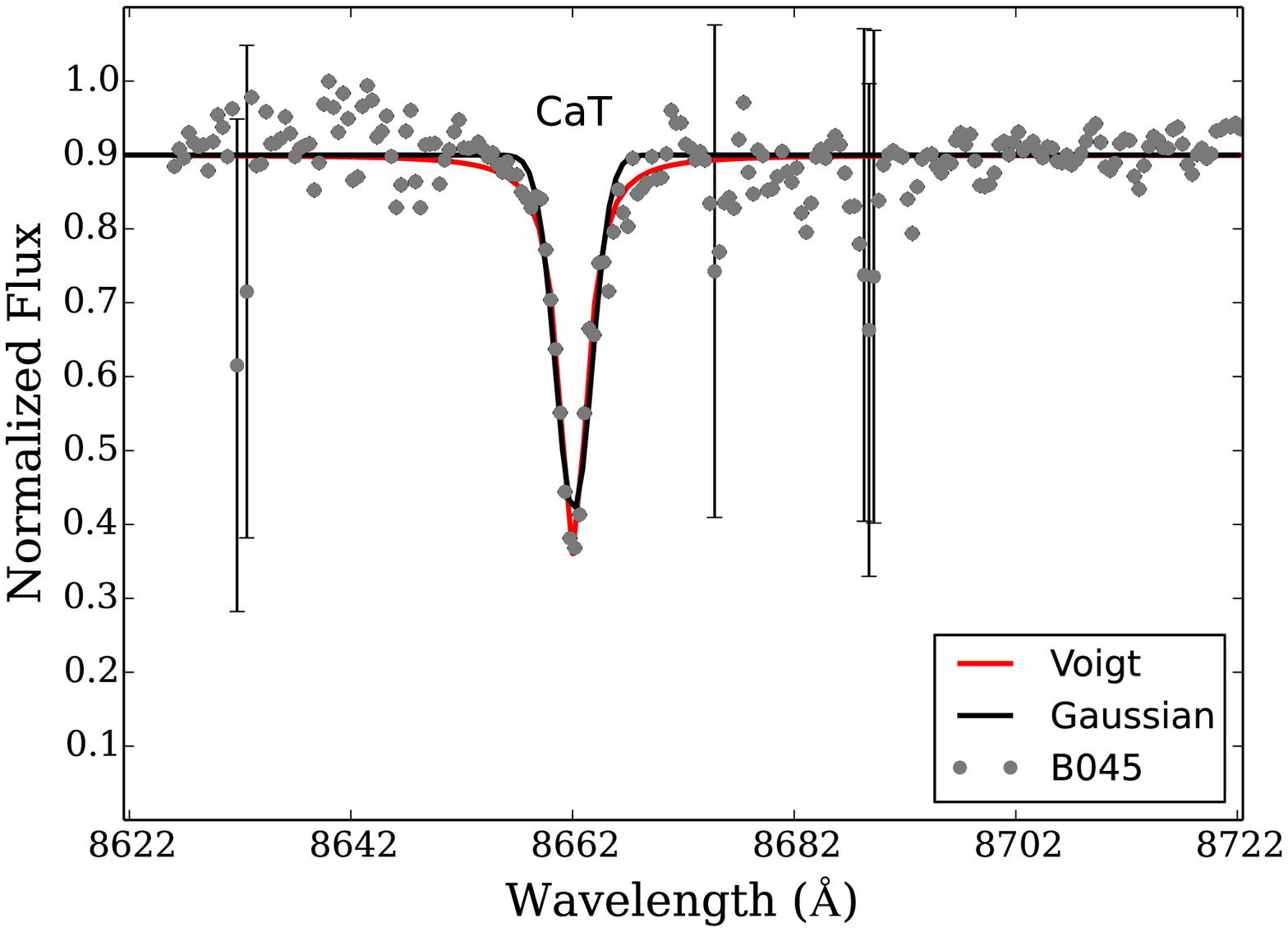}\label{fig:B045CaT1}}
\subfigure{\includegraphics[scale=0.45,trim=0.5in 0in 1.25in 0.0in]{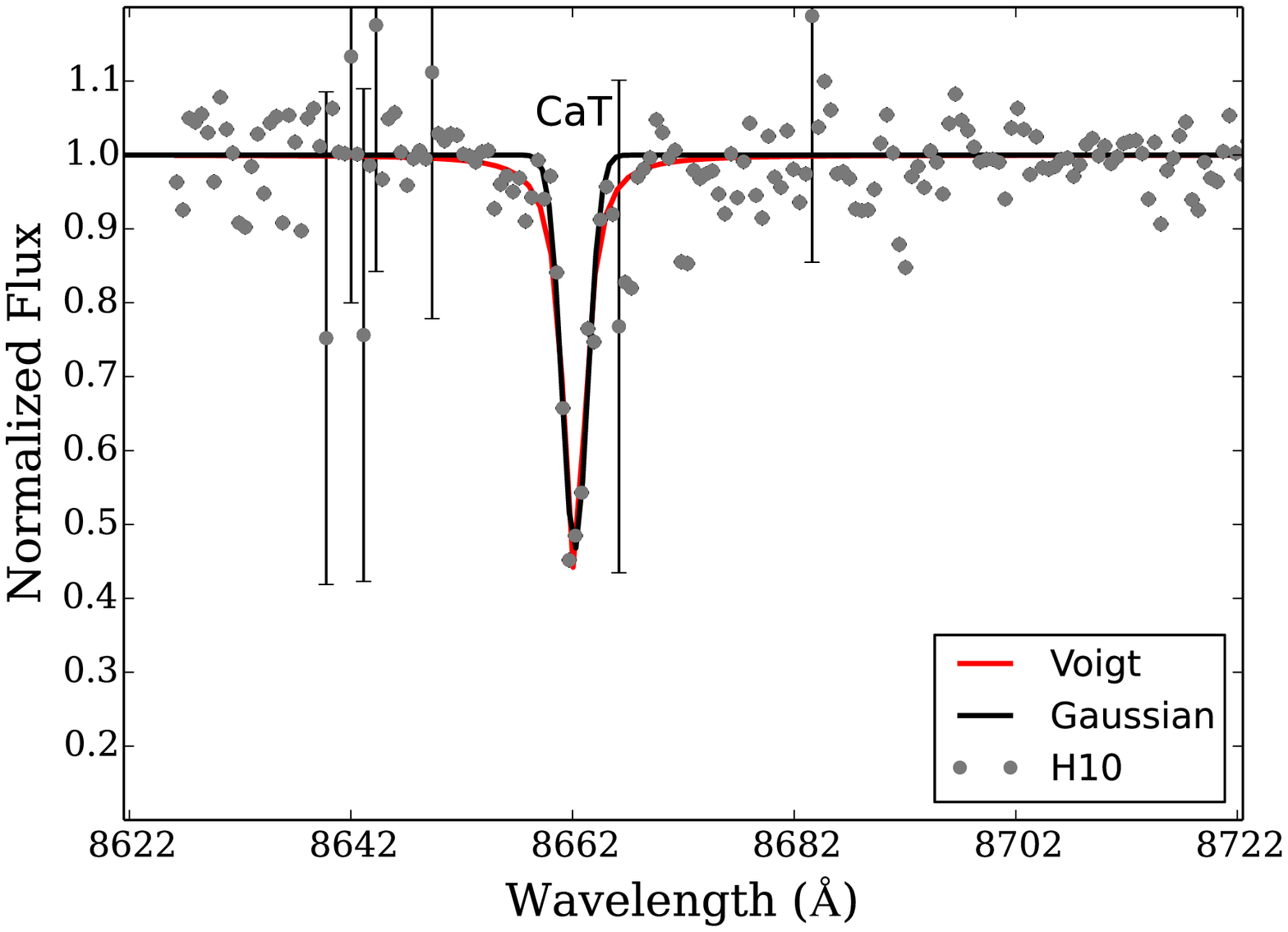}\label{fig:H10CaT3}}
\caption{Sample Gaussian and Voigt profile fits to the third CaT line
  (at 8662 \AA) in B045 (left) and H10 (right).  The grey points show
  the data while the black and red lines show the Gaussian and Voigt
  profile fits, respectively. H10 has lower S/N and a greater
  contamination from sky emission lines; as a result H10's EW
  uncertainties are higher than B045's.\label{fig:SampleFits}}
\end{center}
\end{figure*}

There are a variety of automated line measurement programs.  This
analysis utilizes the {\tt pymodelfit}
program\footnote{\url{https://pythonhosted.org/PyModelFit/}} to
measure Gaussian and Voigt profiles.  Insufficiently removed sky lines
are assigned low weights to prevent them from affecting the line
profiles.  Sample fits to the third CaT line are shown in Figure
\ref{fig:SampleFits} for GCs with moderate and low S/N and some amount
of skyline contamination.  Errors in the fits and resulting line
strengths are determined by resampling the spectra 100 times with
Monte Carlo sampling and bootstrapping.  Note that the continuum level
is fit during the original profile measurement with sky lines and
atomic lines masked out (see Section \ref{subsec:Bandpass}), but is
not remeasured during the resampling.

\begin{figure*}
\begin{center}
\centering
\subfigure{\includegraphics[scale=0.55,trim=1.25in 0in 0.05in 0.0in]{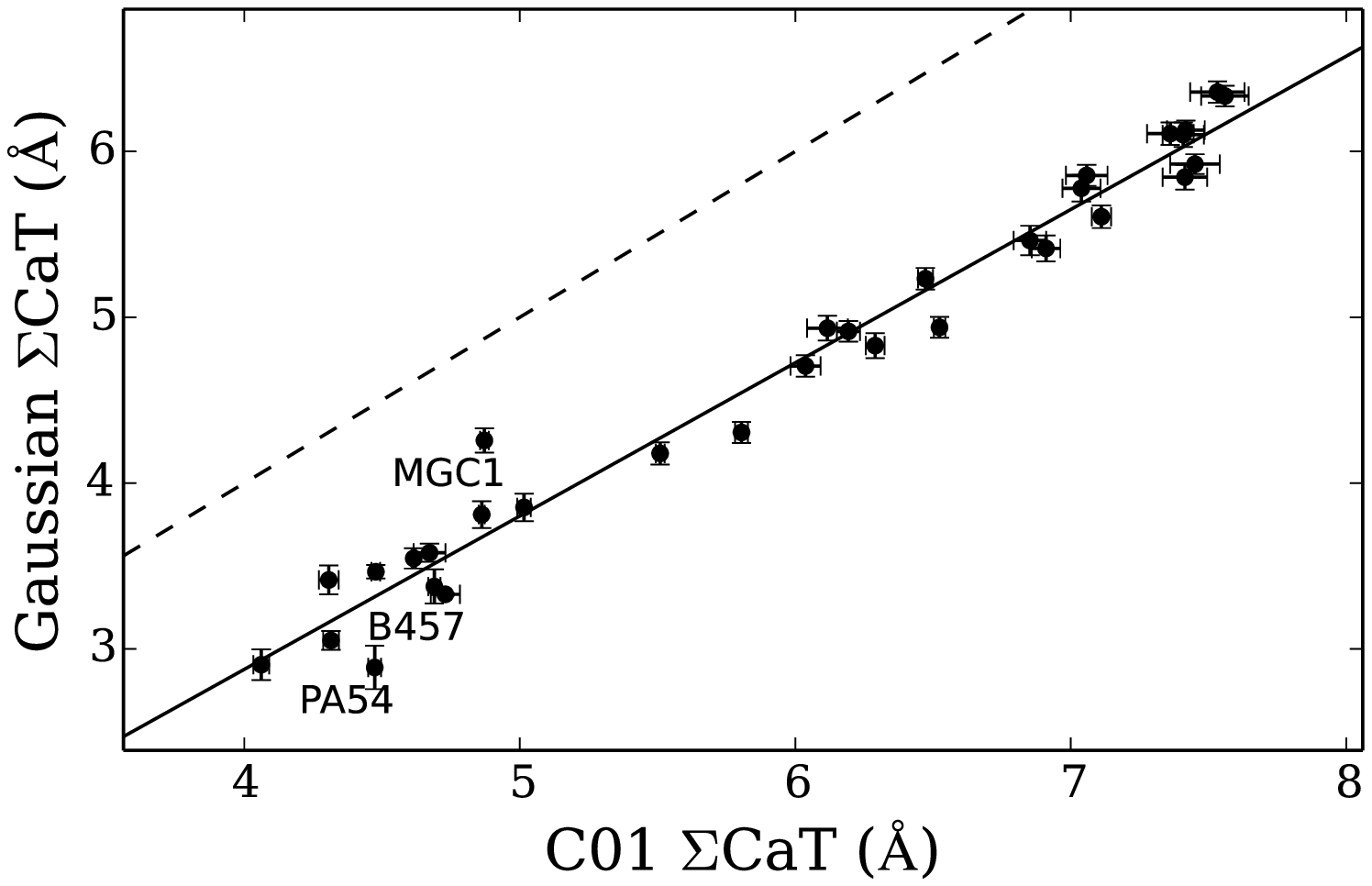}}
\subfigure{\includegraphics[scale=0.55,trim=0.5in 0in 1.25in 0.0in]{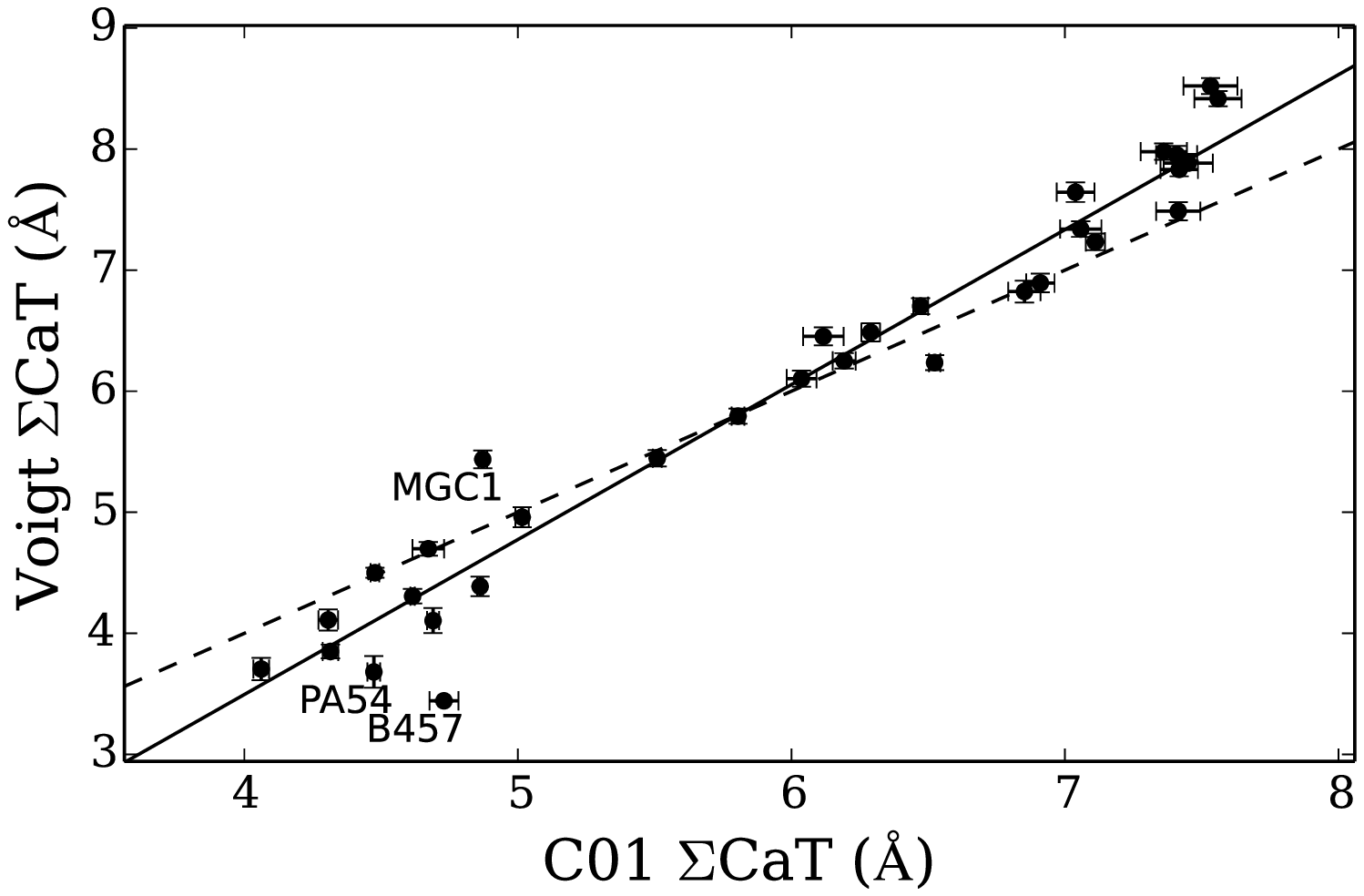}}
\caption{Comparisons between measurements of the IL CaT strength in
  the observed (not template-fitted) spectra.  The points shows
  individual GCs and the $1\sigma$ errors.  The dashed line shows
  perfect agreement, while the solid line shows a linear least squares
  fit.  {\it Left: } Measurements with the C01 definitions (see Section
  \ref{subsubsec:Integrals}) versus Gaussian fits to the three CaT
  lines.
  {\it Right:} Measurements with C01 definitions versus Voigt fits.
  \label{fig:GaussiansVoigts}}
\end{center}
\end{figure*}

Comparisons with the pseudo EWs from Section \ref{subsubsec:Integrals}
are shown in Figure \ref{fig:GaussiansVoigts}.  Significant outliers
are labelled; these are the lower S/N GCs with strong sky
contamination that has affected the pseudo EW measurements.  It is
apparent that the Gaussian fits underpredict the strengths of the CaT,
likely because the strong Lorentzian wings are not included.  The
Voigt profiles match the pseudo EWs for weak CaT strengths; as the CaT
lines strengthen, however, the Voigt profiles become stronger than the
pseudo EWs.  This may be partly due to continuum offsets (see Figure
\ref{fig:C01CompVoigt}).  The scatter is still relatively large
because of noise and residual sky lines, and a comparison with the
template-fitted spectra may be more appropriate.

\begin{figure*}
\begin{center}
\centering
\subfigure{\includegraphics[scale=0.55,trim=1.25in 0in 0.05in 0.0in]{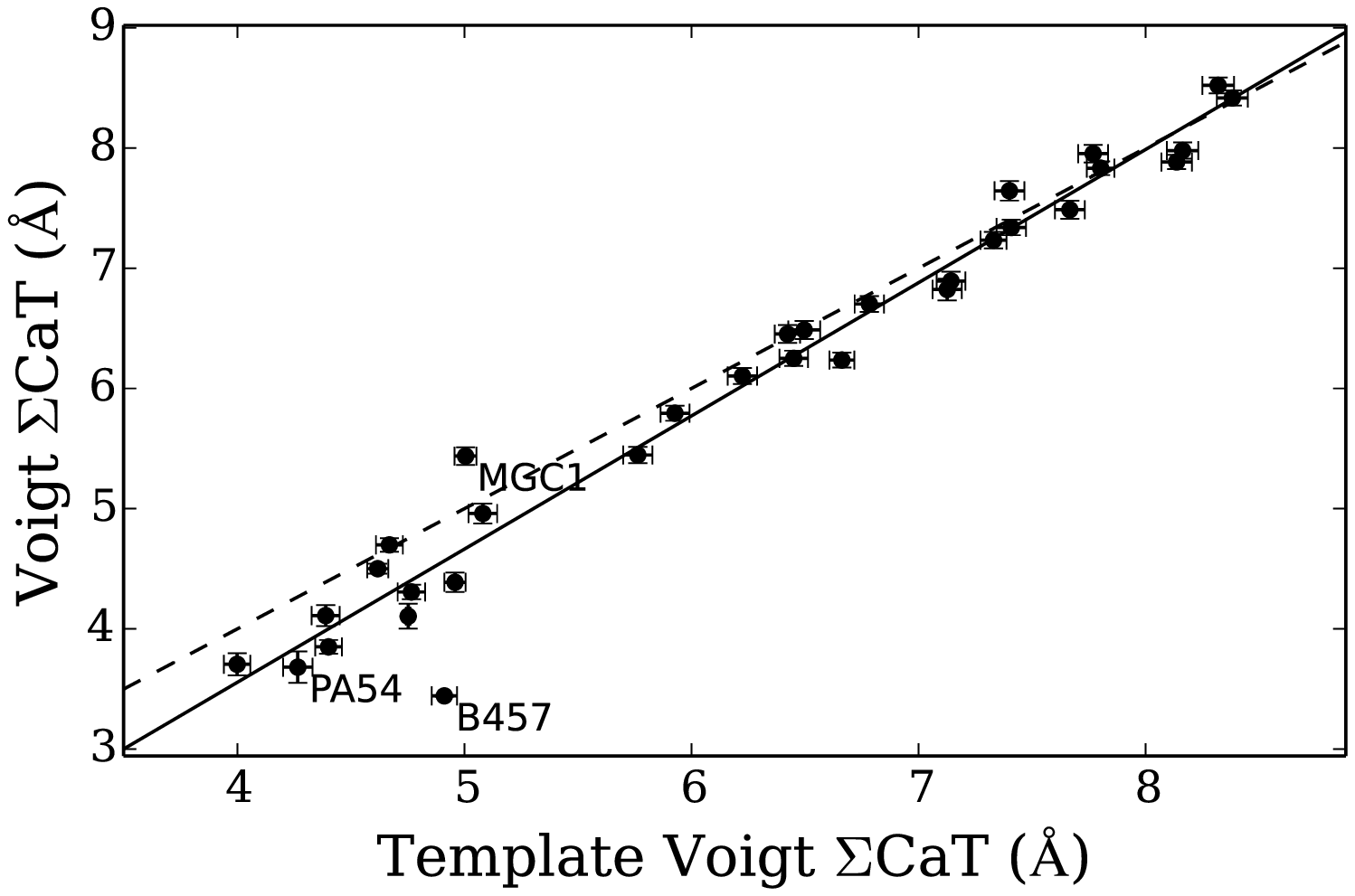}\label{fig:TempVoigtcomp}}
\subfigure{\includegraphics[scale=0.55,trim=0.5in 0in 1.25in 0.0in]{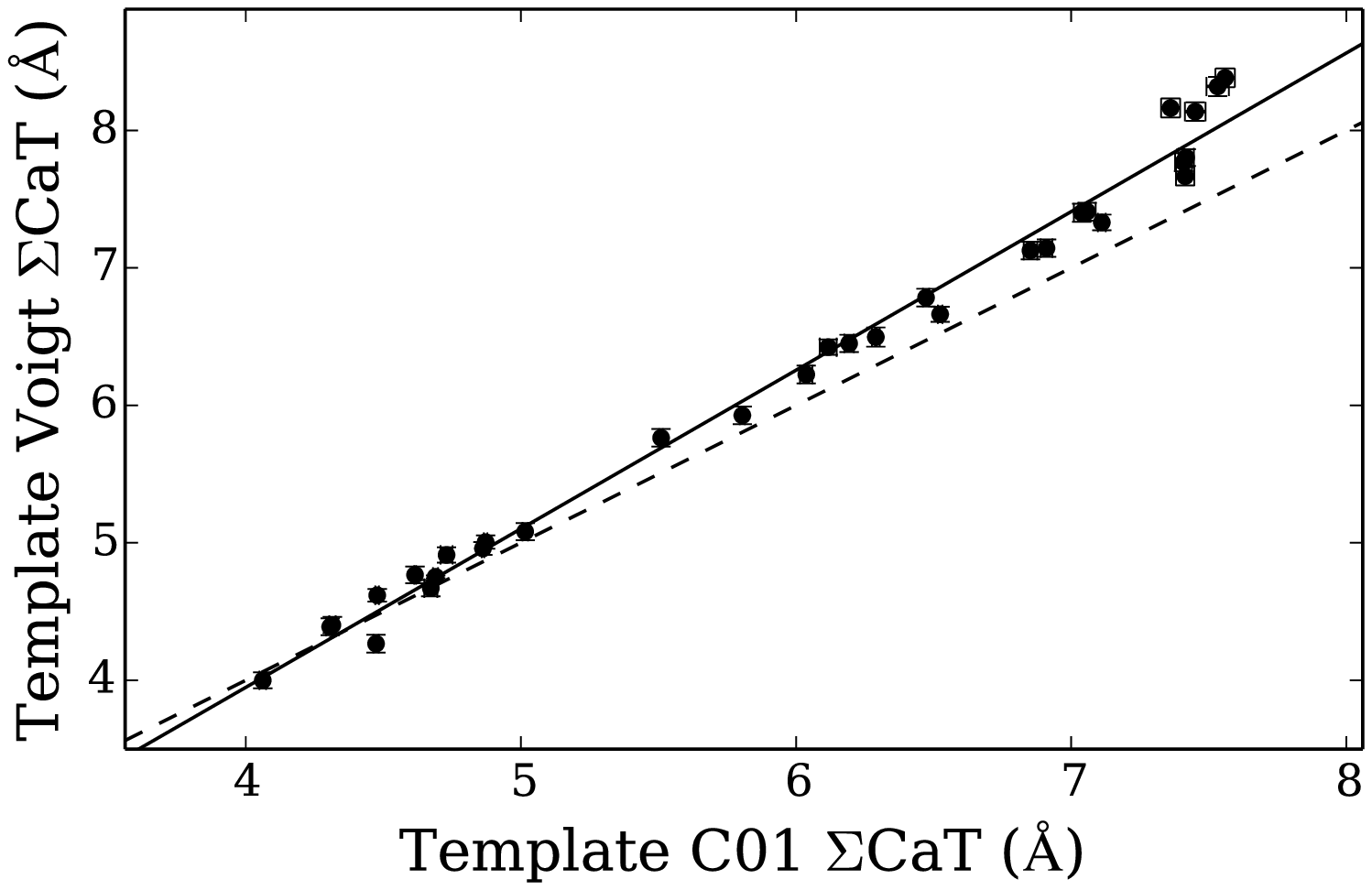}\label{fig:C01comp}}
\caption{Comparisons between measurements of the IL CaT strength in
  the template spectra.  Points are as in Figure \ref{fig:GaussiansVoigts}. {\it Left: } Voigt fits on the raw spectra versus the template spectra.
  {\it Right:} Voigt fits on the template spectra versus pseudo EWs on
  the template spectra with the \citet{Cenarro2001} definitions.
  \label{fig:TemplateComp}}
\end{center}
\end{figure*}

Figure \ref{fig:TempVoigtcomp} compares the Voigt profile fits on the
observed and template-fitted spectra, demonstrating that the agreement
is generally good, with the exception of the low S/N GCs with strong
sky line contamination.  In the latter case, full template fits may
outperform fits of individual lines in the observed spectra.  Figure
\ref{fig:C01comp} then compares Voigt profile fits to integrations
with the C01 indices, both on the template-fitted spectra.  There is a
slight offset that increases with CaT strength---again, this may be
partly due to continuum effects (Figure \ref{fig:C01CompVoigt}), but
is primarily driven by differences in the line measurements.

\section{CaT strength and Metallicity}\label{sec:Metallicity}
Section \ref{sec:Measurements} demonstrated that there are two ideal
ways to accurately measure IL CaT strengths over a wide metallicity
range: 1) integrations of line profiles in template-fitted spectra
with the C01 line and continuum definitions and 2) Voigt profile fits
to observed or template-fitted spectra (though note that Voigt profile
fits on observed spectra may not be ideal for distant, fainter
clusters).  With these measurements the relationship between total CaT
strength and [Fe/H] can be determined.  Because of the disagreement in
radial velocity (see Section \ref{subsec:DataReduction}), B457 has
been removed from these fits.\footnote{Note that B457's CaT strength
  indicates a lower [Fe/H] than the one derived by
  \citet{Colucci2014}, by 0.45 dex; this further
  suggests that different objects were observed.}

\begin{figure*}
\begin{center}
\centering
\subfigure{\includegraphics[scale=0.55,trim=1.25in 0in 0.05in 0.0in]{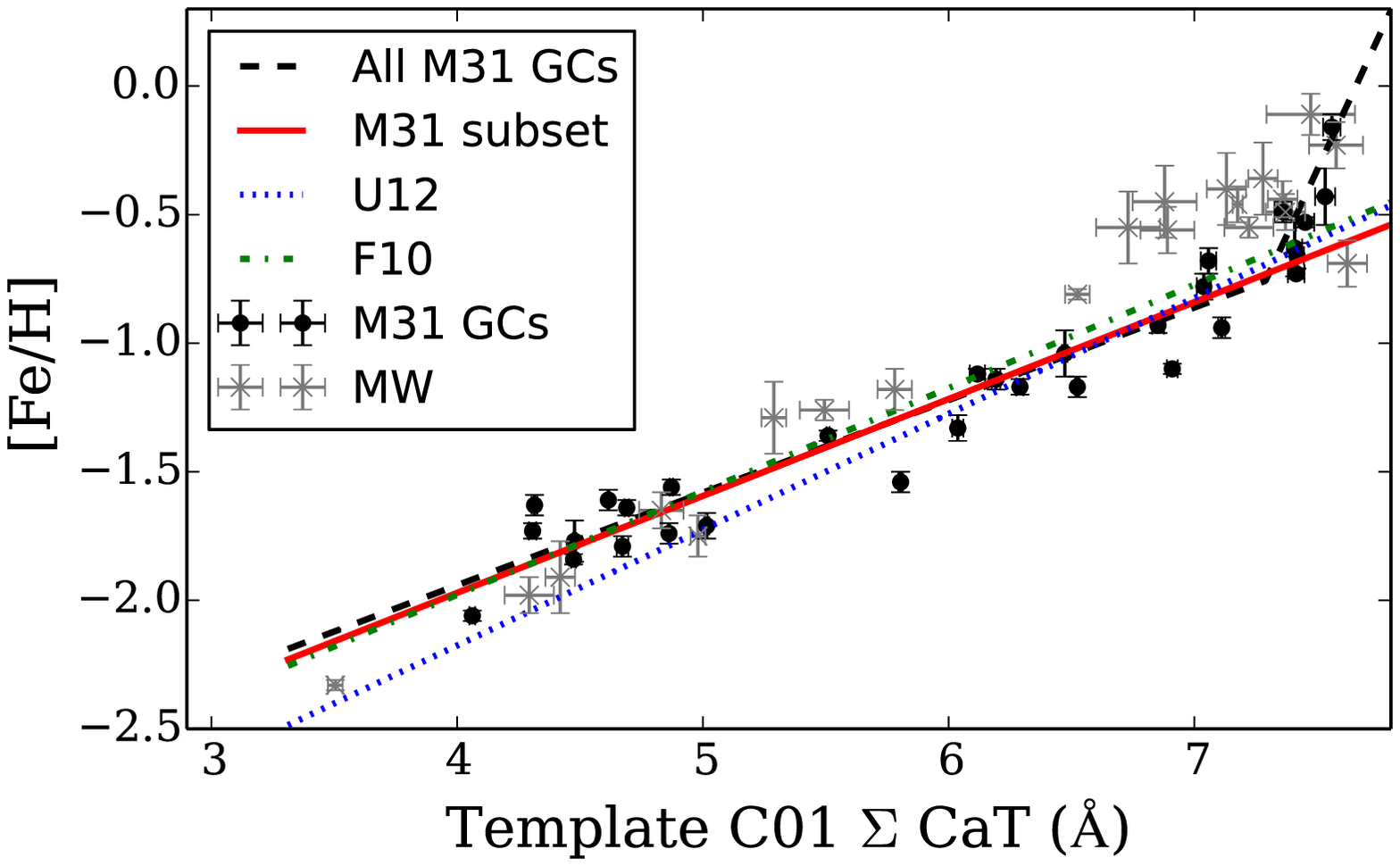}\label{fig:FeHC01}}
\subfigure{\includegraphics[scale=0.55,trim=0.5in 0in 1.25in 0.0in]{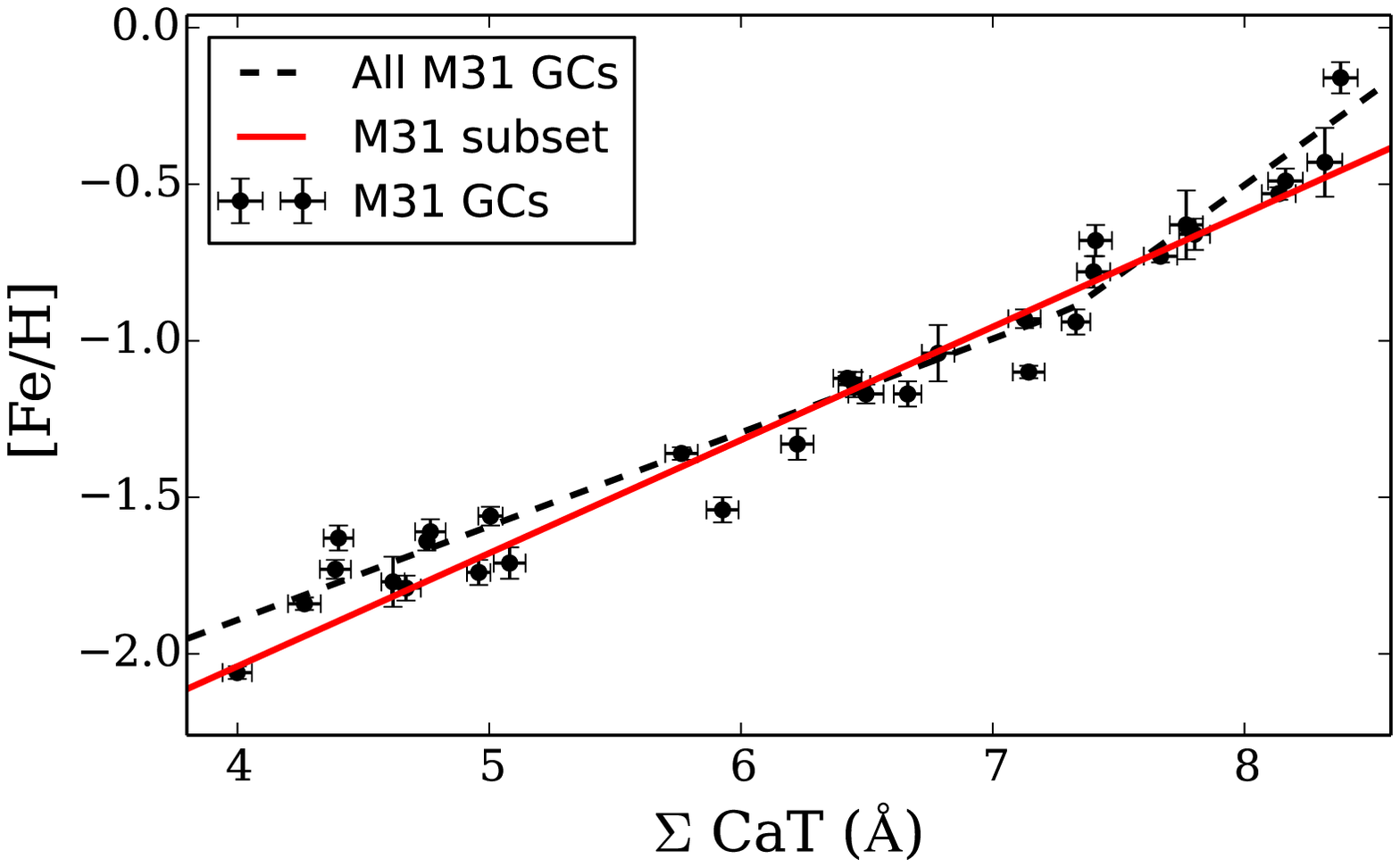}\label{fig:FeHVoigt}}
\caption{High resolution [Fe/H] versus integrated CaT
    strength in the template-fitted spectra, utilizing the C01
    definitions (left) and Voigt profile fits (right).  The black
  circles show the M31 clusters.  The dashed black lines show the
  piecewise linear fits to all M31 GCs; the solid red lines show fits
  without the metal rich GC B193 (see Section \ref{subsec:Breakpoint})
  or the low [Ca/Fe] GCs (see Section \ref{subsec:AlphaFe}).  The grey
  crosses show the MW clusters from AZ88.  The green dot-dashed line
  shows the fit from \citet{Foster2010}, while the dotted blue line
  shows the \citet{Usher2012} fit (after converting from [Z/H] to
  [Fe/H]; see Section \ref{subsec:Othercomp})}.
  \label{fig:FeH}
\end{center}
\end{figure*}

The high resolution [Fe/H] ratios\footnote{The high resolution
  [\ion{Fe}{1}/H] ratios were chosen to represent GC [Fe/H].
  \ion{Fe}{1} has smaller random and systematic errors than
  \ion{Fe}{2} \citep{Sakari2014}, though note that non Local
  Thermodynamic Equilibrium effects and/or uncertainties in modelling
  the underlying stellar populations can lead to disagreements between
  \ion{Fe}{1} and \ion{Fe}{2}.} versus CaT strengths are shown in
Figure \ref{fig:FeH}.  The relationship between CaT strength and
[Fe/H] was assumed to follow a piecewise linear function, in order to
account for possible changes in the slope (see Section
\ref{subsec:Breakpoint}). The optimal slopes, intercepts, and the
breakpoint were found with the SciPy \citep{SciPyREF}
optimize.curve\_fit() package.\footnote{\url{http://www.scipy.org}}
These piecewise linear fits were initialized with a least-squares
linear fit to the full dataset, and with an assumed breakpoint at
$[\rm{Fe/H}] \sim -0.8$.  The fits are mildly sensitive to the initial
slope, but are less sensitive to the initial breakpoint.  Note that
the relationship for Voigt profile fits is very similar between the
observed and template-fitted spectra. Two separate fits were done for
each measurement technique: one utilizing all GCs and another without
the most metal rich GC B193 or the metal poor, low [$\alpha$/Fe]
clusters (G002, MGC1, PA17, PA53; see Sections \ref{subsec:Breakpoint}
and \ref{subsec:AlphaFe}).

\subsection{Breakpoint in the CaT-Metallicity Relationship}\label{subsec:Breakpoint}
Both measurement techniques show a breakpoint in their CaT-[Fe/H]
relations when the most metal cluster, B193, is included in the fit.
This breakpoint has been seen in other IL CaT
studies. \citet{Vazdekis2003} found evidence of a breakpoint in their
models of the CaT, while \citet{Foster2010} saw evidence of a turnover
in their empirical calibration with NGC~1407 GCs.  These studies
referred to this turnover as a ``saturation'' point, which is a
misnomer since the CaT lines are saturated in all the target GCs (that
is, the lines lie on the nonlinear part of the curve of growth).  In
measurements of the CaT strengths in GCs associated with early type
galaxies, \citet{Usher2012} find no evidence for a breakpoint. They
attribute this to their continuum fits, which rely on areas that are
relatively free of weaker atomic lines (much like the continuum
regions defined by C01; see Section \ref{subsec:Bandpass}).  The lack
of a breakpoint for the C01 pseudo EWs and the Voigt profiles (when
B193 is removed) supports the idea that the breakpoint is caused by
line blanketing in the continuum regions.  Since neither continuum
fitting technique can adequately fit B193, this suggests that the CaT
strength-[Fe/H] may become uncertain above $[\rm{Fe/H}] \sim -0.4$.

\begin{figure}
\begin{center}
\centering
\hspace{-0.35in}
\includegraphics[scale=0.52]{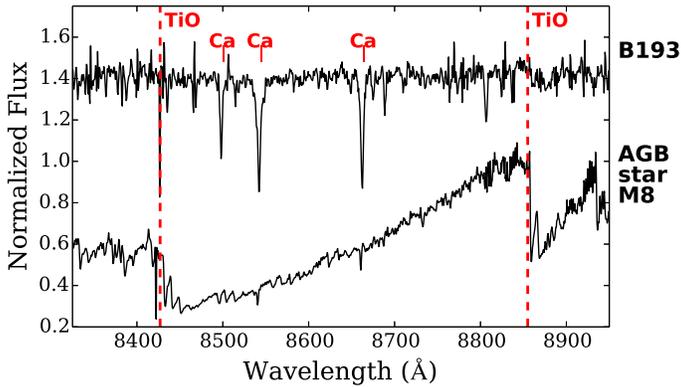}
\caption{The IL CaT spectrum of the metal rich GC B193, along with a
  spectrum of an M8 AGB star.  The cool AGB star's spectrum is
  dominated by TiO bands; the two strongest bandheads at 8432 and 8860
  \AA \hspace{0.025in} are indicated with red dashed lines.  The 8860
  \AA \hspace{0.025in} feature is also detectable in the B193
  spectrum, which hints that the CaT features may also be affected by
  TiO absorption from M giants in the cluster.
  \label{fig:TiO}}
\end{center}
\end{figure}

It is important to note that continuum fits become increasingly
difficult as a GC's metallicity increases.  Not only do atomic lines
become stronger, molecular lines become increasingly important as
well.  Metal rich GCs are also likely to host cool M giants with
strong TiO absorption in the CaT region; in IL spectra M giants may
have a non-negligible contribution to the CaT region.  Their molecular
lines will likely affect the continuum {\it and} the lines themselves,
and may be difficult to detect.  The CaT lines in B193 ($[\rm{Fe/H}]
\sim -0.2$) may be affected by molecular line blanketing, leading to
weaker measured CaT lines with all measurement techniques.  This is
demonstrated in Figure \ref{fig:TiO}, which shows the CaT spectra of
B193 and an M8 metal rich AGB star.  The red dashed lines show the
strongest TiO bandheads (also see Figure 1 in C01).  B193 seems to
have some moderate absorption at 8860 \AA, which implies that there
will be weak TiO absorption throughout the CaT region.  TiO absorption
was also seen in the stacked red (metal rich) spectra from
\citet{Usher2015}.  The numbers of M giants in metal rich GCs (and
their precise temperatures) is likely to be somewhat stochastic; as a
result, TiO blanketing may lead to scatter in the [Fe/H]-CaT relation
at high [Fe/H].

\subsection{The Effects of Detailed Chemical Abundances}\label{subsec:AlphaFe}
The CaT lines are sensitive to the electron pressure in stellar
atmospheres; nonstandard chemical abundance mixtures may affect the
measured CaT strength and the inferred [Fe/H].  A handful of GCs in
this sample are chemically distinct from typical MW and M31
GCs---that is they have abundance ratios that indicate they may have
formed in dwarf galaxies that were later accreted by M31.  One of the
most telling signs of accretion is a low [Ca/Fe] ratio at a given
[Fe/H].  In addition, several of these chemically anomalous GCs are
located on or near stellar streams, with radial velocities that link
them to the streams and/or to each other \citep{Veljanoski2014}. The
clusters that are likely associated with stellar streams include H10,
PA53, and PA56, while the chemically peculiar GCs (in more than one
abundance ratio) without a known stream or host galaxy include G002,
MGC1, and PA17.

\begin{figure}
\begin{center}
\centering
\hspace{-0.2in}
\includegraphics[scale=0.55,trim=0.4in 0in 0in 0.0in]{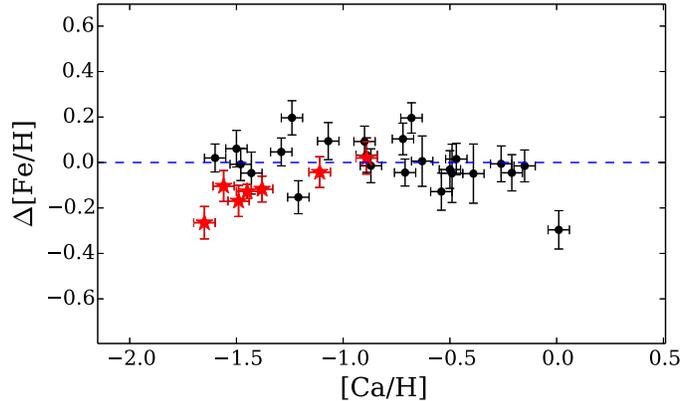}
\caption{Offsets from the best fitting CaT strength-metallicity
  relationship versus [Ca/H] when the chemically peculiar GCs are
  excluded from the fit.  The GCs with low [Ca/Fe], other abundance
  ratios, and/or locations along stellar streams that indicate they
  may have formed in dwarf galaxies are indicated with red stars;
  these GCs are G002, MGC1, H10, PA17, PA53, and PA56.
  \label{fig:DiffCaFe}}
\end{center}
\end{figure}

Figure \ref{fig:DiffCaFe} shows the offsets from the CaT
strength-metallicity relation for the Voigt profile fits to
template-fitted spectra versus [Ca/H], with the chemically interesting
GCs indicated with red stars. [Ca/H] is utilized rather than [Ca/Fe],
since the metal rich GCs also have lower [Ca/Fe] ratios (due to
standard chemical evolution; see \citealt{Tolstoy2009}).  To separate
chemical abundance offsets from continuum issues, [Ca/H] is used
instead of [Ca/Fe].  B193 clearly sticks out at the metal rich end
(see Section \ref{subsec:Breakpoint}).  Only the fits with the low
[Ca/Fe] GCs (and B193) excluded from the fit is shown.  Though
there is no definite trend in Figure \ref{fig:DiffCaFe}, the metal
poor, chemically peculiar GCs stand out, with their CaT strengths
systematically lower than the ``normal'' GCs. This indicates that a
GC's detailed abundances can affect its CaT line strengths,
particularly at the metal poor end.  For this reason, the best CaT
strength-metallicity relations are determined without these GCs.

\subsection{Comparisons with Previous Results}\label{subsec:Othercomp}
Empirical relationships between integrated CaT strength and GC
metallicity have been determined in three major studies: the original
Galactic GC calibration by AZ88, the recalibration of the AZ88
relation by \citet{Foster2010}, and the SSP-based calibration of
\citet{Usher2012}.  These papers all used slightly different methods
to determine CaT strengths, and are discussed individually.

\subsubsection{AZ88's Original Calibration: Milky Way GCs}
The CaT strengths from AZ88 and the high resolution [Fe/H] ratios can
be directly compared to those of the M31 GCs.  The MW CaT lines were
measured with the AZ88 line definitions and continuum regions.
Sections \ref{subsec:Bandpass} and \ref{subsubsec:Integrals}
demonstrated that there is a clear offset between pseudo EWs measured
with the AZ88 vs. C01 definitions.  This offset has been applied to
adjust the MW CaT strengths, and the MW GCs are shown in Figure
\ref{fig:FeHC01}.  The adopted metallicities for the
Galactic GCs are from \citet{Harris}, to ensure that the MW GCs
are on the same [Fe/H] scale; the Harris metallicities are from
compilations of high resolution analyses of individual stars, and
should roughly match the high resolution metallicities from
\citet{Colucci2014} and \citet[2016 {\it in prep.}]{Sakari2015}.  For
47~Tuc and M15 the IL [Fe/H] ratios from \citet{Sakari2013} are
adopted.

The MW GCs are slightly offset from the M31 GCs, particularly at
intermediate metallicities.  This may be due to metallicity
offsets. However, it may also be an effect of spectral resolution.
The AZ88 measurements were made on lower resolution spectra
($R\sim2000$); the resulting blending may make continuum levels even
more difficult to identify.  As a result, the breakpoint in the
relation may come at a lower metallicity than the M31 GCs.  Regardless
of the cause, the MW GCs generally agree with the M31 GCs.

\subsubsection{The \citet{Foster2010} Rederivation of AZ88}
\citet{Foster2010} rederived the AZ88 CaT strength-[Fe/H] relationship
with the same line definitions but with better continuum fitting,
ensuring better performance at high metallicity.   The Foster et
al. relation was shifted to the wider C01 line definition (see Section
\ref{subsubsec:Integrals}), and is also shown in Figure
\ref{fig:FeHC01}.  The agreement with the M31 GC relation is
excellent; the relations start to diverge slightly at the metal rich
end, though the difference is not significant.

\subsubsection{The \citet{Usher2012} SSP-based Relation}
\citet{Usher2012} present two CaT-metallicity calibrations: a
rederivation of the Foster et al. relation and a new calibration based
on SSP models.  As Usher et al. select the latter relation as the
better choice, only the SSP-based model is considered here.  As with
\citet{Foster2010}, Usher et al. use the AZ88 line definition but
perform more robust continuum identifications; their relation has
therefore also been shifted to the C01 line definitions (see Section
\ref{subsubsec:Integrals}).  However, their relation is given in [Z/H]
rather than [Fe/H].  It is difficult to know how to compare
theoretical total metallicities to spectroscopic [Fe/H] ratios.  In
principle there are two approaches.
\begin{enumerate}
\item Convert GC [Fe/H] ratios to [Z/H] using detailed abundances.  In
  practice this approach is extremely difficult, since large
  star-to-star variations exist within GCs, in some of the most
  abundant elements (e.g. He, C, N, O, Na, Mg, Al) due to evolutionary
  and/or GC processes
  (e.g. \citealt{Gratton2000,Briley2004,Carretta2009}).  This is
  further complicated by potentially large systematic errors which may
  plague observed abundance ratios \citep{Sakari2014}.  Finally, the
  SSP models do not account for light element variations, and this is
  likely to not be a robust way to compare the two relations.
\item Convert SSP [Z/H] ratios to [Fe/H].  In principle this approach
  should be simpler than converting observationally-determined
  abundance ratios, since the inputs to the model are known.  However,
  the precise abundance mixture and the metallicity scale will affect
  the final [Z/H].  This is the method that is adopted for this
  comparison.
\end{enumerate}

Usher et al. adopted solar-scaled SSPs and then adjusted those values
to account for varying [$\alpha$/Fe], following the prescription of
\citet{Mendel2007}.\footnote{Note that \citet{Mendel2007} adopt the
  conversion from \citet{Trager2000}, assuming an $\alpha$-enhanced,
  N-enhanced, solar C profile for all GCs.  Since a GC's IL is
  dominated by tip of the RGB stars, the assumption of high N and
  solar C is reasonable; infrared IL observations of these
  M31 GCs confirm this result (Sakari et al. 2016, {\it in prep.}).}
To compare with the M31 GCs, these [Z/H] ratios were then converted
to the Harris et al. (1996, 2003 edition) utilizing Usher et al.'s
Equation 3.  Technically, their Equation 3 was determined for a
different set of SSP models than was used for their final CaT
strength-metallicity relationship; however, their [$\alpha$/Fe]
correction should correct for any offset.  The converted relationship
is shown in Figure \ref{fig:FeHC01}.

The metallicity scale is very important for this conversion.  Usher et
al. also provide a calibration on the \citet{CarrettaFe} scale, which
leads to a more discrepant slope from the M31 GCs.  Undoing the
\citet{Mendel2007} correction leads to a large offset, which indicates
that some adjustment to a metallicity scale is necessary.  As with the
AZ88 GCs, the Harris metallicity scale is chosen because its [Fe/H]
ratios come primarily from high resolution, individual stellar
spectroscopy, and should provide the best agreement with the M31
literature values.  Even with the Harris metallicity scale, there is
some evidence for a slope difference.  This could indicate further
problems in the metallicity scale, or it could indicate problems with
the Usher et al. scale at low [Fe/H].
 
\subsection{Paschen Lines}\label{subsec:Paschen}
The template-fitted M31 spectra of the metal poor GCs show slight
Paschen features.  \citet{Usher2015} also detected weak Paschen lines
in their stacked spectra of blue (metal poor) GCs.  These weak Paschen
lines do not have a significant affect on the CaT strengths in these
GCs, which are all older than $\sim 2$ Gyr.  However, they may be
significant in younger clusters.  C01 define a CaT* index that
accounts for nearby Paschen lines. This index seems to overcorrect the
CaT strengths in these old M31 GCs, but may be essential for young
clusters.

\subsection{Versions of CaT Strength}\label{subsec:Combination}
Various studies have utilized different combinations of the CaT lines.
For future studies of more distant clusters, it may be desirable to
utilize different combinations of lines; for instance, the first CaT
line may be too weak to detect confidently or the third CaT line may
be completely obscured by sky lines.  For this reason, Table
\ref{table:FeH} presents the CaT strength-metallicity relationships
for different combinations of the CaT lines.  Three different
combinations are considered: the classic sum of all three CaT lines,
the sum of the strongest two (the second and third;
e.g. \citealt{Tolstoy2001}), and the second CaT line by itself.
Naturally the relation for the second CaT line is more uncertain than
the others.  The slope of the CaT2$+$CaT3 relation is also slightly
steeper than the relation with all three lines.

Note that when Paschen contamination is significant (e.g. for younger
GCs), the first CaT line may be more reliable for measuring GC
metallicity \citep{Wallerstein2012}.  Given that the first CaT line is
weaker, the errors on this relationship are larger, and this relation
is not included in Table \ref{table:FeH}.

Given that the derived relations are very sensitive to the measurement
techniques, the M31 CaT spectra from this paper are available online.\footnote{\url{http://faculty.washington.edu/sakaricm/CaT.html}}

\begin{table*}
\centering
\begin{minipage}{165mm}
\begin{center}
\caption{CaT strength vs. metallicity relationships, for different
  indicators of the CaT strength.\label{table:FeH}}
  \begin{tabular}{@{}llc@{}}
  \hline
& & \\
Measurement & CaT Strength & Relation   \\
Technique   & Indicator    &            \\
\hline
& & \\
Pseudo EWs (C01 definitions)& $\Sigma$CaT & $[\rm{Fe/H}] = (0.38\pm 0.10)\times\rm{CaT} - (3.48\pm 0.13)$ \\
 & CaT2 + CaT3 & $[\rm{Fe/H}] = (0.45\pm 0.10)\times \rm{CaT} -
(3.45\pm 0.20)$ \\
 & CaT2 & $[\rm{Fe/H}] = (0.78\pm 0.20)\times \rm{CaT} - (3.36\pm 0.33)$ \\
& & \\
Voigt Profile Fits & $\Sigma$CaT & $[\rm{Fe/H}] = (0.36\pm 0.10)\times \rm{CaT} - (3.49\pm 0.20)$ \\
 & CaT2 + CaT3 & $[\rm{Fe/H}] = (0.44\pm 0.20)\times \rm{CaT} - (3.49\pm 0.19)$ \\
 & CaT2 & $[\rm{Fe/H}] = (0.74\pm 0.40)\times \rm{CaT} - (3.32\pm 0.50)$ \\
\hline
\end{tabular}
\end{center}
\end{minipage}\\
\medskip
\end{table*}

\subsection{Age Effects}\label{subsec:Age}
\begin{figure}
\begin{center}
\centering
\includegraphics[scale=0.55]{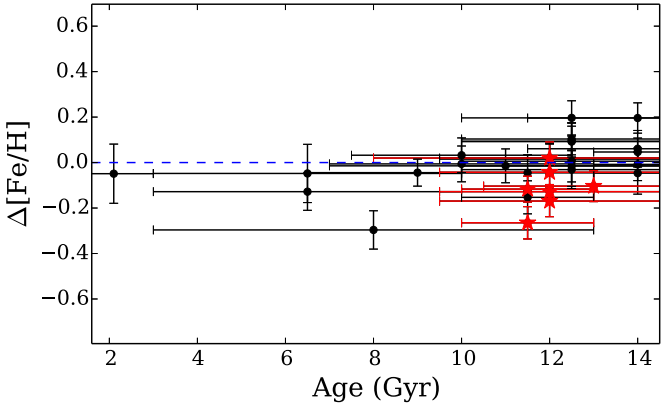}
\caption{Offsets from the best fitting CaT strength-metallicity
  relationship versus age.  Points are as in Figure
  \ref{fig:DiffCaFe}.  There is no significant relationship between GC
  age and offset from the best-fitting CaT strength-metallicity
  relation.
  \label{fig:DiffAge}}
\end{center}
\end{figure}

Figure \ref{fig:DiffAge} shows offsets from the CaT
strength-metallicity relationship versus GC age, where the ages are
from the high resolution spectroscopic analyses of
\citet{Colucci2014} and \citet[2016 {\it in prep.}]{Sakari2015}.  Down
to $\sim 2$ Gyr, there is no significant trend, suggesting that age
has little effect on the IL CaT (although the large error bars on the
age prohibit any subtle trend from being detected).  This may not be
true for younger GCs if strong Paschen lines begin to contaminate the
CaT lines and continuum regions; in that case the C01 CaT* index may
help to remove Paschen contamination.

\section{Discussion: Interpretations of CaT Strengths}\label{sec:Discussion}
Considered as a whole, the tests with these well-studied, relatively
nearby M31 GCs have reproduced the indirect calibrations of
\citet{Foster2010} and \citet{Usher2012}, at least for the observed
metallicity range.  However, these tests have also shown that CaT
measurements are not trivial, particularly for metal rich GCs with
significant line blanketing.  This line blanketing makes it difficult
to identify continuum levels, and will become even worse with large
velocity dispersions.  Nonstandard chemical abundance mixtures
(e.g. for GCs that appear to have formed in dwarf galaxies) also
confuse the CaT relation at the low metallicity
end. \citet{Foster2010} and \citet{Usher2012} found several intriguing
inconsistencies between CaT and colour-based metallicities, which they
have attributed to continuum problems and age or chemical abundance
differences.  With this set of nearby M31 GCs, which have been well
studied, these hypotheses can start to be tested.

\subsection{Metallicity Bimodalities}\label{subsec:Bimodalties}
In their study of NGC~1407, \citet{Foster2010} found that the general
shape of the CaT bimodality (which was only detected in
template-fitted spectra) did not match the shape of the colour
bimodality.  \citet{Foster2011} also noted an excess of GCs in
NGC~4494 at $[\rm{Fe/H}] = -0.9$ (determined with their CaT
calibration), which corresponds to a trough in the colour
distribution.  B193's large offset from the best-fitting relation
suggests that Foster et al.'s distribution was affected by line
blanketing in the most metal rich GCs, which led to continuum issues
and subsequent weakening of the measured CaT lines.  If the
best-fitting relation in this paper were applied to GCs at B193's
metallicity, they would also seem to pile up at lower metallicities,
weakening any intrinsic bimodality with a significant metal rich
population.  Though this problem may be alleviated with more rigorous
continuum fits, it is likely to remain a difficulty for metal rich
GCs.

\subsection{The Blue Tilt}\label{subsec:BlueTilt}
The blue tilt is a photometric phenomenon where the brightest GCs in
the blue (metal poor) subpopulation appear redder than their fainter
counterparts.  There has also been weak photometric evidence for
inverse red tilts (\citealt{Harris2009}) and positive red tilts
\citep{Mieske2010}, although neither case is very significant.  These
tilts have been attributed to abundance variations within the most
massive GCs, a hypothesis that is borne out by observations of some MW
GCs, e.g. NGC~1851 (e.g. \citealt{Carretta2010}), NGC~3201 (Hughes et
al. 2015, {\it in prep.}), and $\omega$ Cen
(e.g. \citealt{Johnson2010}), which have large spreads in many
elements, including heavy elements like Fe. Even monometallic GCs like
M15 show large star-to-star variations in Mg, which could affect its
integrated colours (e.g. \citealt{Sneden1997}).

\citet{Foster2010} found that the brightest GCs in both subpopulations
have identical CaT strengths, leading to identical CaT-based
metallicities despite being well separated in colour.
\citet{Usher2012} confirmed this result.  Possible explanations for
this behaviour included non universal colour-CaT strength relations
and chemical or age differences between GC populations.  However, it
is also possible that the CaT strengths of the brightest metal rich
GCs could be systematically underpredicted because of continuum
uncertainties.  If the brighter GCs are more massive, they will have
higher velocity dispersions, leading to more blending of the atomic
features---such an effect would also be seen at lower spectral
resolution (e.g. with the MW GCs; see Section \ref{subsec:Othercomp}).
C01 and \citet{Vazdekis2003} explored velocity dispersion and
resolution effects and concluded that the strengths of the C01 indices
were unaffected by broadening, up to large velocity dispersions.
However, the continuum levels are likely to be affected, which may be
difficult to detect in low S/N spectra.  Similarly, narrow line
definitions will be affected by line broadening; while the C01 indices
are unaffected, the AZ88 definitions may be affected by velocity
dispersion.  Continuum offsets at a fixed metallicity with increasing
cluster mass would indeed be detectable as a negative tilt in the
metal rich GCs. \citet{Usher2015} find no evidence for a red tilt in
their stacked spectra.

Thus, while the blue tilt could indeed be a result of multiple
populations within GCs, a red tilt could be created (or exaggerated)
by systematic offsets between massive and low mass GCs at the same
metallicity.  This could also lead to observational biases when
observing faint GC systems where only the most massive GCs are
observable.

The tests in Section \ref{subsec:AlphaFe} also demonstrated that the
detailed abundance mixture does affect the CaT strength and the
inferred metallicity.  None of the chemically peculiar GCs
investigated here are massive enough to create a blue tilt, but the
most massive, metal poor GCs are likely to have strong star-to-star
variations in elements that could complicate interpretations of the
CaT strength (e.g. Mg; \citealt{Carretta2009}).

\subsection{CaT Metallicity-Colour Relationships}\label{subsec:CaTColor}
The most intriguing result from the \citet{Usher2012} analysis is that
the CaT metallicity-colour relationship does not seem to be the same
for all galaxies.  In particular, while there is excellent agreement
between GCs at intermediate metallicities in all galaxies, certain
galaxies have large discrepancies at high or low metallicity.  By
stacking spectra from GCs with similar colours, \citet{Usher2015}
confirmed these CaT-colour differences between early-type galaxies.
These differences are driven mostly at the metal rich end, though at
the metal poor end differences are seen between galaxies with luminous
vs. faint GC systems.  Most of the GCs are quite bright, and are 
therefore likely to be massive.

The results from this paper suggest that care should be taken at the
metal rich end.  Although \citet{Usher2015} argue that there is no
evidence for a red tilt in their stacked spectra (i.e. the CaT
strength does not depend on GC magnitude), stochastic effects
(e.g. the numbers of cool M giants) in metal rich GC populations could
lead to a spread in the CaT strength-metallicity relationship.  Fe
spreads in the most massive GCs could also confuse CaT-colour
relationships.

At the metal poor end, detailed chemical abundances can affect the
slope of the CaT-[Fe/H] relationship.  GCs that formed in dwarf
galaxies and have primordial abundance differences from MW and M31
stars and GCs have lower predicted [Fe/H] ratios at a given CaT
strength.  While it is unlikely that the early type galaxies host
significant populations of dwarf galaxy clusters,  they almost
certainly host variations in He, C, N, O, Na, Mg, and/or
Al. \citet{Usher2015} propose that these variations would lead to
colour differences, but they may also lead to differences in CaT
strength.  With these chemical variations it is difficult to
interpret a GC's total metallicity.

The takeaway message from this M31 GC analysis is therefore one of
caution: CaT-based metallicities are highly sensitive to measurement
techniques, and are likely to become increasingly difficult with
higher [Fe/H] and velocity dispersion.

\section{Conclusion}\label{sec:Conclusion}
This paper has presented the first comparison between integrated CaT
measurements and high resolution integrated [Fe/H] since the original
\citet{AZ88} analysis.  The M31 GCs used for this analysis span a wide
metallicity range, from $[\rm{Fe/H}]~=~-2$ to $-0.2$, and cover ages
from $\sim 2-14$ Gyr.  The results from Sections
\ref{sec:Measurements} and \ref{sec:Metallicity} demonstrate several
crucial aspects of the CaT as a metallicity indicator.
\begin{enumerate} 
\item The quantitative strength of the CaT lines depends on the
  measurement techniques.  Voigt profiles best fit the lines, though
  line integrations also work, provided that the line bandpasses are
  defined appropriately.
\item Template fits to the observed spectra do not introduce
  significant systematic offsets in the measured CaT strengths, 
  provided that the templates cover a sufficient range in parameter
  space.
\item The CaT line strengths are significantly affected by continuum
  fits.  Continuum fits that rely on continuum bandpasses (such as the
  AZ88 definitions) will be blanketed by atomic lines in GCs with
  $[\rm{Fe/H}]\ga -1$.  The most metal rich GCs may have strong
  molecular features that are contributed by cool M giants; this
  contamination may be seen in the TiO bandhead at 8860 \AA.  {\it It
    is extremely difficult to fit continuum levels properly in these
    most metal rich clusters.}  Similarly, high velocity dispersion
  GCs will have atomic lines blended together, further complicating
  continuum estimates.
\item If continuum levels are properly fit the integrated CaT is an
  excellent [Fe/H] indicator, to within $\sim 0.2$ dex.  The precise
  relationship depends mildly on the techniques used to the measure
  the line strengths and on the specific lines used.
\item Age does not affect the CaT line strengths in any of these
  clusters, which are older than $\sim 2$ Gyr.
\item Detailed abundance mixtures may also play a small role in the
  derived [Fe/H].
\end{enumerate}
Interpretations of CaT-based metallicities must consider the
difficulties in accurately measuring CaT strength, the potential
biases that may occur from only observing the brightest GCs, and any
additional systematic effects that could alter the integrated CaT
strength.

\section*{Acknowledgments}
The authors would like to thank the anonymous referee for suggestions
that improved this manuscript.  The authors also thank Dianne Harmer
and Joanne Hughes for their assistance with the WIYN telescope
observations, and the observing specialists at APO and KPNO for their
assistance and expertise.  CMS and GW acknowledge funding from the
Kenilworth Foundation.  This research has made use of the SIMBAD
database, operated at CDS, Strasbourg, France.


\footnotesize{

}


\begin{thebibliography}{99}
\bibitem[Armandroff \& Zinn(1988)] {AZ88} Armandroff, T.E. \& Zinn,
  R. 1988, \aj, 96, 92
\bibitem[Battaglia et al.(2008)] {Battaglia2008} Battaglia, G., Irwin,
  M., Tolstoy, E., et al. 2008, \mnras, 383, 183
\bibitem[Bershady et al.(2004)] {Bershady1} Bershady, M.A., Andersen,
  D.R., Harker, J., Ramsey, L.W., \& Verheijen, M.A.W. 2004, \pasp,
  116, 565
\bibitem[Bershady et al.(2005)] {Bershady2} Bershady, M.A., Andersen,
D.R., Verheijen, M.A.W., Westfall, K.B., Crawford, S.M., \& Swaters,
R.A. 2005, \apjs, 156, 311
\bibitem[Bershady et al.(2008)] {Bench1} Bershady, M., Barden, S.,
  Blanche, P.-A., et al. 2008, \spie, 7014
\bibitem[Briley et al.(2004)] {Briley2004} Briley, M.M., Harbeck, D.,
  Smith, G.H., \& Grebel, E.K. 2004, \aj, 127, 1588
\bibitem[Brodie \& Strader(2006)] {BrodieStrader2006} Brodie, J.P. \&
  Strader, J. 2006, \araa, 44, 193
\bibitem[Caldwell et al.(2011)] {Caldwell2011} Caldwell, N., Schiavon,
R., Morrison, H., Rose, J.A., \& Harding, P. 2011, \aj, 141, 18
\bibitem[Cappellari \& Emsellem(2004)] {pPXFref} Cappellari, M. \&
  Emsellem, E. 2004, \pasp, 116, 138
\bibitem[Cardiel(2010)] {Cardiel2010} Cardiel, N. 2010, {\it
  Astrophysics Source Code Library}
\bibitem[Carretta et al.(2009a)] {Carretta2009} Carretta, E.,
Bragaglia, A., Gratton, R., \& Lucatello, S. 2009a, \aap, 505, 139
\bibitem[Carretta et al.(2009b)] {CarrettaFe} Carretta, E., Bragaglia,
A., Gratton, R., D'Orazi, V., \& Lucatello, S. 2009b, \aap, 508, 695
\bibitem[Carretta et al. (2010)] {Carretta2010} Carretta, E.,
  Gratton, R.G., Lucatello, S., et al. 2010, \apjl, 722, L1
\bibitem[Cenarro et al.(2001)] {Cenarro2001} Cenarro, A.J., Cardiel,
  N., Gorgas, J., Peletier, R.F., Vazdekis, A., \& Prada, F. 2001,
  \mnras, 326, 959
\bibitem[Colucci et al.(2009)] {Colucci2009} Colucci, J.E., Bernstein,
R.A., Cameron, S., McWilliam, A., \& Cohen, J.G. 2009, \apj, 704, 385
\bibitem[Colucci et al.(2011a)] {Colucci2011a} Colucci, J.E.,
Bernstein, R.A., Cameron, S.A., \& McWilliam, A. 2011, \apj, 735, 55
\bibitem[Colucci et al.(2011b)] {Colucci2011b} Colucci, J.E. \&
Bernstein, R.A. 2011, EAS Pub. Ser., 48, 275
\bibitem[Colucci et al.(2012)] {Colucci2012} Colucci, J.E., Bernstein,
R.A., Cameron, S.A., \& McWilliam, A. 2012, \apj, 746, 29
\bibitem[Colucci et al.(2013)] {Colucci2013} Colucci, J.E., Duran,
M.F., Bernstein, R.A., \& McWilliam, A. 2013, \apjl, 773, 36
\bibitem[Colucci et al.(2014)] {Colucci2014} Colucci, J.E., Bernstein,
  R.A., \& Cohen, J.G. 2014, \apj, 797, 116
\bibitem[Conroy \& van Dokkum(2012)] {ConroyVanDokkum2012} Conroy,
  C. \& van Dokkum, P.G. 2012, \apj, 760, 71
\bibitem[Ferreras et al.(2013)] {Ferreras2013} Ferreras, I., La
  Barbera, F., de la Rosa, I.G., et al. 2013, \mnras, 429, 15
\bibitem[Foster et al.(2010)] {Foster2010} Foster, C., Forbes, D.A.,
  Proctor, R.N., Strader, J., Brodie, J.P., \& Spitler, L.R. 2010,
  \aj, 139, 1566
\bibitem[Foster et al.(2011)] {Foster2011} Foster, C., Spitler, L.R.,
  Romanowsky, A.J., et al. 2011, \mnras, 415, 3393
\bibitem[Galleti et al.(2004)] {RBCref} Galleti, S., Federici, L.,
  Bellazzini, M., Fusi Pecci, F., \& Macrina, S. 2004, \aap, 416, 917
\bibitem[Galleti et al.(2006)] {RBCref2} Galleti, S., Federici, L.,
  Bellazzini, M., Buzzoni, A., \& Fusi Pecci, F. 2006, \aap, 456, 985 
\bibitem[Gratton et al.(2000)] {Gratton2000} Gratton, R.G., Sneden,
  C., Carretta, E., \& Bragaglia, A. 2000, \aap, 354, 169
\bibitem[Harris(1996, 2010 edition)] {Harris} Harris, W.E. 1996 (2010
  edition), \aj, 112, 1487
\bibitem[Harris(2009)] {Harris2009} Harris, W.E. 2009, \apj, 699, 254
\bibitem[Harris et al.(2013)] {Harris2013} Harris, W.E., Harris,
  G.L.H., \& Alessi, M. 2013, \apj, 772, 82
\bibitem[Hinkle(2003)] {Hinkle2003} Hinkle, K., Wallace, L.,
Livingston, W., Ayres, T., Harmer, D., \& Valenti, J. 2003, 	
in \textit{The Future of Cool-Star Astrophysics: 12th Cambridge
Workshop on Cool Stars, Stellar Systems, and the Sun (2001 July 30 -
August 3)}, eds. A. Brown, G.M. Harper, and T.R. Ayres, (University of
Colorado), 851
\bibitem[Huxor et al.(2014)] {Huxor2014} Huxor, A.P., Mackey, A.D.,
Ferguson, A.M.N., et al. 2014, \mnras, 442, 2165
\bibitem[Johnson \& Pilachowski(2010)] {Johnson2010} Johnson, C.I. \&
  Pilachowski, C.A. 2010, \apj, 722, 1373
\bibitem[Jones et al.(2001)] {SciPyREF} Jones, E., Oliphant, E.,
  Peterson, P., et al. 2001, \url{http://www.scipy.org}
\bibitem[Knezek et al.(2010)] {Bench2} Knezek, P.M., Bershady, M.A.,
  Willmarth, D., et al. 2010, \spie, 7735
\bibitem[McWilliam \& Bernstein(2008)] {McWB} McWilliam, A. \&
Bernstein, R. 2008, \apj, 684, 326
\bibitem[Mendel et al.(2007)] {Mendel2007} Mendel, J.T., Proctor,
  R.N., \& Forbes, D.A. 2007, \mnras, 379, 1618
\bibitem[Mieske et al.(2010)] {Mieske2010} Mieske, S., Jord\'{a}n, A.,
  C\^{o}t\'{e}, P., et al. 2010, \apj, 710, 1672
\bibitem[Minkowski(1942)] {Minkowski1942} Minkowski, R. 1942, \apj,
  96, 306
\bibitem[Peng et al.(2006)] {Peng2006} Peng, E.W., Jord\'{a}n, A.,
  C\^{o}t\'{e}, P. et al. 2006, \apj, 639, 95
\bibitem[Sakari et al.(2013)] {Sakari2013} Sakari, C.M., Shetrone, M.,
Venn, K., McWilliam, A., \& Dotter, A. 2013, \mnras, 434, 358
\bibitem[Sakari et al.(2014)] {Sakari2014} Sakari, C.M., Venn, K.,
Shetrone, M., Dotter, A., \& Mackey, D. 2014, \mnras, 443, 2285
\bibitem[Sakari et al.(2015)] {Sakari2015} Sakari, C.M., Venn, K.A.,
  Mackey, A.D., Shetrone, M.D., Dotter, A., Ferguson, A.M.N., \&
  Huxor, A. 2015, \mnras, 448, 1314
\bibitem[Sneden et al.(1997)] {Sneden1997} Sneden, C., Kraft, R.P.,
  Shetrone, M.D., Smith, G.H., Langer, G.E., \& Prosser, C.F. 1997,
  \aj, 114, 1964
\bibitem[Starkenburg et al.(2010)] {Starkenburg2010} Starkenburg, E.,
  Hill, V., Tolstoy, E., et al. 2010, \aap, 513, 34
\bibitem[Tolstoy et al.(2001)] {Tolstoy2001} Tolstoy, E., Irwin, M.J.,
  Cole, A.A., Pasquini, L., Gilmozzi, R., \& Gallagher, J.J. 2001,
  \mnras, 327, 918
\bibitem[Tolstoy et al.(2009)] {Tolstoy2009} Tolstoy, E., Hill, V., \&
Tosi, M. 2009, \araa, 47, 371
\bibitem[Trager et al.(2000)] {Trager2000} Trager, S.C., Faber, S.M.,
  Worthey, G., \& Gonz\'{a}lez, J.J. 2000, \aj, 119, 1645
\bibitem[Usher et al.(2012)] {Usher2012} Usher, C., Forbes, D.A.,
  Brodie, J.P., et al. 2012, \mnras, 426, 1475
\bibitem[Usher et al.(2015)] {Usher2015} Usher, C., Forbes, D.A.,
  Brodie, J.P., et al. 2015, \mnras, 446, 369
\bibitem[Vazdekis et al.(2003)] {Vazdekis2003} Vazdekis, A., Cenarro,
  A.J., Gorgas, J., Cardiel, N., \& Peletier, R.F. 2003, \mnras, 340,
  1317
\bibitem[Veljanoski et al.(2014)] {Veljanoski2014} Veljanoski, J.,
Mackey, A.D., Ferguson, A.M.N., et al. 2014, \mnras, 442, 2929
\bibitem[Wallerstein et al.(2012)] {Wallerstein2012} Wallerstein, G.,
  Gomez, T., \& Huang, W. 2012, {\it ApSS}, 341, 89
\bibitem[Worthey et al.(1994)] {Worthey1994} Worthey, G., Faber,
  S.M., Gonz\'{a}lez, J.J., \& Burstein, D. 1994, \apjs, 94, 687
\bibitem[Yoon et al.(2006)] {Yoon2006} Yoon, S.-J., Yi, S.K., \& Lee,
  Y.-W. 2006, {\it Science}, 311, 1129
\end{thebibliography}
\end{document}